\documentclass[aps,reprint,amsmath,amssymb,groupaddress,superscriptaddress,pra]{revtex4-2}
\usepackage{graphicx}
\usepackage[utf8]{inputenc}
\usepackage{amsmath}
\usepackage{subfigure}
\usepackage{amsfonts}
\usepackage{amssymb}
\usepackage{bm}
\usepackage{dcolumn}
\usepackage{array}  
\usepackage{diagbox}
\usepackage{color}
\usepackage[utf8]{inputenc}
\usepackage[T1]{fontenc}
\usepackage{booktabs, array, mathptmx, float, tabularx, booktabs, lipsum, amsmath, multirow,mathtools}
\graphicspath{{figs/}{figsgaoerb/}}
\usepackage{soul}
\usepackage{xcolor}
\usepackage{mathtools}
\usepackage{lipsum}
\usepackage{comment}

\usepackage{xcolor}
\usepackage[colorlinks,citecolor=blue,linkcolor=blue,anchorcolor=blue,urlcolor=blue]{hyperref}

\begin{document}

\title{Unlocking inaccessible performance of the quantum refrigerator with catalysts}

\author{Cong Fu}
\affiliation{Department of Physics, Xiamen University, Xiamen 361005, Fujian, China}
\affiliation{Department of Physics, Fujian Provincial Key Laboratory for Soft Functional Materials Research,\\Xiamen University, Xiamen 361005, China}

\author{Ousi Pan}
\affiliation{Department of Physics, Xiamen University, Xiamen 361005, Fujian, China}

\author{Zhiqiang Fan}
\affiliation{Department of Physics, Xiamen University, Xiamen 361005, Fujian, China}

\author{Yushun Tang}
\affiliation{Department of Physics, Xiamen University, Xiamen 361005, Fujian, China}

\author{Shanhe Su}
\email{sushanhe@xmu.edu.cn}
\affiliation{Department of Physics, Xiamen University, Xiamen 361005, Fujian, China}
\affiliation{Department of Physics, Fujian Provincial Key Laboratory for Soft Functional Materials Research,\\Xiamen University, Xiamen 361005, China}

\author{Youhui Lin}
\email{linyouhui@xmu.edu.cn}
\affiliation{Department of Physics, Xiamen University, Xiamen 361005, Fujian, China}
\affiliation{Department of Physics, Fujian Provincial Key Laboratory for Soft Functional Materials Research,\\Xiamen University, Xiamen 361005, China}

\author{Jincan Chen}
\email{jcchen@xmu.edu.cn}
\affiliation{Department of Physics, Xiamen University, Xiamen 361005, Fujian, China}

\date{\today}
\begin{abstract}
Quantum thermal machines offer promising platforms for exploring the fundamental limits of thermodynamics at the microscopic scale. The previous study in Ref.~\cite{lobejko2024,biswas2024} demonstrated that the incorporation of a catalyst can significantly enhance the performance of a heat engine by broadening its operational regime and achieving a more favorable trade-off between work output and efficiency. Building on this powerful framework and innovative idea, here we further extend the concept to a two-stroke quantum refrigerator that extracts heat from a cold reservoir via discrete strokes powered by external work. The working medium consists of two two-level systems (TLSs) and two heat reservoirs at different temperatures and is assisted by an auxiliary system acting as a catalyst. Remarkably, the catalyst remains unchanged after each cycle, ensuring that heat extraction is driven entirely by the work input. We show that the presence of the catalyst leads to two significant enhancements: it enables the coefficient of performance (COP) and cooling capacity to exceed the Otto bound and allows the refrigerator to operate in frequency and temperature regimes that are inaccessible without a catalyst. Furthermore, through a comparison with catalytic heat engines, our analysis reveals that two distinct permutation types are necessary to simultaneously enhance the COP and operational range of refrigerators, in contrast to heat engines for which a single permutation suffices. These results highlight the potential of catalytic mechanisms to broaden the operational capabilities of quantum thermal devices and to surpass conventional thermodynamic performance limits. 
\end{abstract}
\maketitle

\section{Introduction}
\label{I}

Inspired by its role in chemical processes, the catalyst in quantum thermodynamics~\cite{allahverdyan2011,aberg2014,brand2015,ng2015,lostaglio2015,sparaciari2017,wilming2017,muller2018,boes2020,shiraishi2021,lipka-bartosik2021a,czartowski2024,junior2024,deoliveirajunior2025} and information~\cite{jonathan1999,daftuar2001,turgut2007,sandersNecessary2009,aberg2014,majenz2017,anshu2018,boes2018,rethinasamy2020,ding2021} is an auxiliary system that facilitates transformations otherwise prohibited by standard thermodynamic laws. Within the framework of quantum thermodynamic resource theory, such a catalyst has been extensively studied for facilitating state interconversions under unitary and Gibbs-preserving transformations. In particular, the concept of catalytic majorization,~\cite{daftuar2001,turgut2007}, along with its extensions~\cite{muller2018}, provides a rigorous framework to quantify how a catalyst can reshape the occupancy distribution of a working medium, thereby enabling state transitions that go beyond standard thermodynamic bounds ~\cite{bhatia1997,rethinasamy2020}.
 
In conjunction with Birkhoff’s theorem~\cite{bhatia1997} and Blackwell’s theorem~\cite{D.Blackwell1953}, the catalyst plays a pivotal role in elucidating the fundamental limits of work extraction from a quantum system. These mechanisms offer critical insights into improving the performance of cyclic quantum heat engines~\cite{henao2021,henao2023,biswas2024,lobejko2024}. By introducing elements such as stochastic independence~\cite{lostaglio2015}, correlations~\cite{shiraishi2021}, and entropy~\cite{Von2019,wilming2021}, the catalyst serves as operational thermodynamic resources. They enable a structured and systematic approach to surpass standard thermodynamic limits through precise control over quantum states~\cite{lobejko2024}.

\begin{figure}		\includegraphics[width=0.48\textwidth]{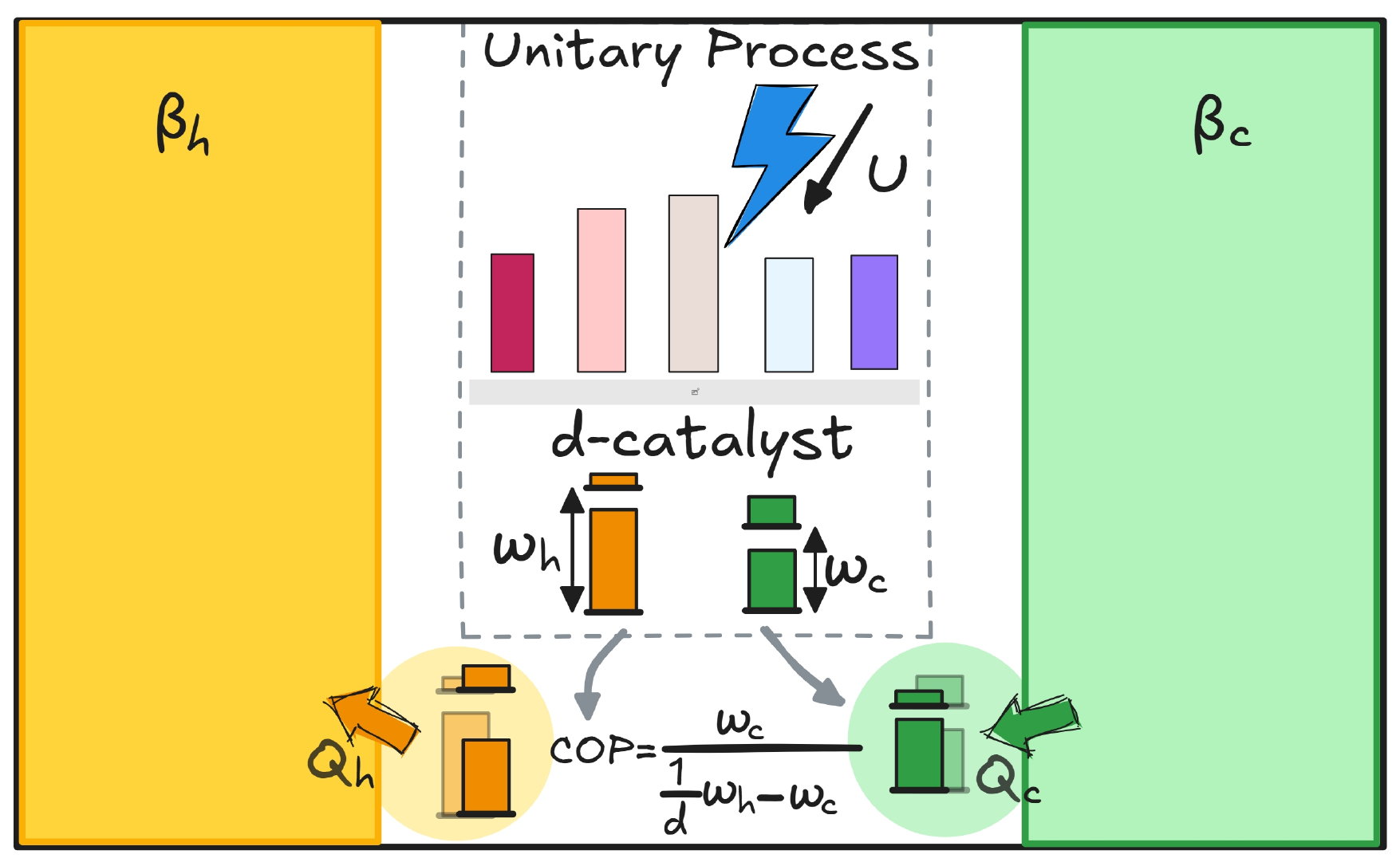}
    \caption{Two-stroke quantum refrigerator assisted by a catalyst. The device is operated to cool a target cold reservoir characterized by inverse temperature $\beta_c$, transferring the extracted energy to a hot environment at inverse temperature $\beta_h$ with the assistance of an external work source $W$ and a catalytic auxiliary system. }
	\label{fig:refrigerator}
    \end{figure}

Beyond heat engines, quantum cooling represents a fundamental task in quantum thermodynamics, enabling the control of heat flow at the microscopic scale~\cite{Brunner2014,clivaz2019, Correa2014}. Extensive prior work has centered on performance optimization of quantum refrigerators under finite-time constraints and exploration of their dynamics across various system–reservoir coupling schemes~\cite{levy2012a,kolar2012,venturelli2013,Correa2014,clivaz2019,soldati2022}. 
Recently, the schemes based on catalytic majorization have been employed to lower the average energy of quantum states, thereby reducing heat dissipation and ultimately achieving efficient cooling~\cite{sparaciari2017,shiraishi2021,henao2023}. Henao and Uzdin showed that correlations emerging during information erasure can be exploited catalytically to suppress heat dissipation, thereby achieving environmental cooling through energy redistribution under controlled unitary dynamics~\cite{henao2021}. In follow-up studies, they showed that finite-dimensional catalysts facilitate cooling in finite-size systems by breaking the passive-state condition via carefully engineered population transfers~\cite{henao2023}. However, these approaches are primarily restricted to non-cyclic transformations~\cite{sparaciari2017}. Their application to fully cyclic thermodynamic refrigerator remains largely unexplored. Notably, while catalysts have been studied for mitigating irreversibility in quantum cycles, their potential for actively enhancing cycle performance remains to be systematically explored. Recently, catalytic techniques have been demonstrated to enhance the performance of microscopic heat engines operating within a two-stroke Otto-like paradigm~\cite{lobejko2024}. In particular, {\L}obejko and Biswas et al showed that introducing a finite-dimensional catalyst into a two-stroke engine yields a generalized "$d$-Otto" efficiency and significantly extends its operational regime~\cite{lobejko2024,biswas2024}. Based on this powerful framework and constructive protocol, we propose the integration of catalytic elements into quantum cooling cycles, offering new insights into the design of practical refrigeration schemes and underscoring the broader role of catalysis in quantum thermodynamics.

In this paper, we fill this gap by presenting a fully characterized model of a catalyst‐assisted, two‐stroke quantum cooling cycle rather than focusing on the generic state interconversion in the thermodynamic scenario~\cite{jonathan1999}. We specify the catalyst state and unitary protocol, and derive closed-form expressions for both the performance metrics and the cooling window. These results reveal that, beyond enhancing the COP and $Q_c$, the presence of the catalyst significantly broadens the operational regime of the quantum refrigerator. Our findings demonstrate that catalytic resources can be effectively leveraged in the concrete design of quantum thermal machines with enhanced performance. The concrete contents of the paper are organized as follows: In Sec.~\ref{II}, we provide a mathematical description of two-stroke refrigerators operating in the microscopic regime and define the key physical quantities required to quantitatively assess their performance. In Sec.~\ref{III}, we present a mathematical framework for calculating the optimal performance of a noncatalytic two-stroke refrigerator by considering all $24$ possible permutations. In Sec.~\ref{IV} and Sec.~\ref{V}, we demonstrate how to construct specific permutation protocols to enhance the optimal COP and expand the operational regime in the presence of a catalyst. In Sec.~\ref{VI}, we present a performance comparison between catalytically enhanced refrigerators and heat engines. Finally, we draw our conclusions in Sec.~\ref{VII}.

\section{Thermodynamics of a two-stroke refrigerator}
\label{II}

We consider a two-stroke quantum refrigerator, schematically shown in Fig.~\ref{fig:refrigerator}. The working medium consists of a $d$-dimensional catalyst with Hamiltonian $H_s = \sum_m m\omega |m\rangle\langle m|$, a TLS with Hamiltonian $H_h = \omega_h |1\rangle\langle 1|$ contacted with a hot reservoir, and a TLS with Hamiltonian $H_c = \omega_c |1\rangle\langle 1|$ contacted with a cold reservoir. Here, the ground-state energies of both TLSs are set to zero for simplicity and without loss of generality. In this notation, $\omega$ represents the frequency difference between any pair of energy levels of the catalyst subsystem, while $\omega_h$ and $\omega_c$ correspond to the energy gaps of the hot and cold TLSs, respectively. The catalyst is in the state $\rho_s=\sum_m p_m |m\rangle\langle m|$, where $p_m$ denotes the probability of corresponding occupying state. It features multiple discrete energy levels, effectively enhancing the systematic dimensionality and providing greater flexibility in energy level permutations. The initial state of the working medium is defined as the product state: $\rho=\tau_h \otimes \tau_c \otimes \rho_s$, where 
\begin{equation}
\tau_h=\frac{e^{-\beta_h H_h}}{\operatorname{Tr}\left(e^{-\beta_h H_h}\right)} ; \quad \tau_c=\frac{e^{-\beta_c H_c}}{\operatorname{Tr}\left(e^{-\beta_c H_c}\right)}.
    \label{tau_h,c}
\end{equation} 
Note that $\beta_h$ and $\beta_c$  are the inverse temperatures of the hot and cold reservoirs, respectively.

The first stroke is realized through a global unitary operation $U$, generated by an externally applied pulse. The state of the working medium changes from $\rho$ to $\rho ^\prime =U\rho U^{\dag}$. This unitary operation enables the redistribution of occupation probabilities among different energy levels, while maintaining the catalytic state invariant over the full cycle. That is, the final marginal state of the catalyst equals its initial state, i.e., $\mathrm{Tr}_{h,c} \left[ U\rho U^{\dag} \right]=\rho_s $. 

During the second stroke, the TLSs disengage from the catalyst and rethermalize with their respective heat reservoirs. The total system recovers the initial state $\rho_s$ and thereby ensuring the cyclic operation of the refrigerator. Specifically, the heat released by the working medium into the hot reservoir is quantified by
\begin{equation}
Q_h=\mathrm{Tr}\left[ H_h(\rho ^\prime-\rho) \right]
    \label{Qh}
\end{equation}
and the heat transferred to the working medium from the cold reservoir is expressed as
\begin{equation}
Q_c=\mathrm{Tr}\left[ H_c(\rho -\rho^\prime) \right].
    \label{Qc}
\end{equation}
According to the first law of thermodynamics, the work supplied by the external agent is 
\begin{equation}
W=Q_h-Q_c=\mathrm{Tr}\left[ (H_h+H_c)(\rho^\prime -\rho) \right].
    \label{W}
\end{equation}
The COP of the refrigerator is given by
\begin{equation}
\mathrm{COP}=\frac{Q_c}{W}=\frac{Q_c}{Q_h-Q_c}.
    \label{COP}
\end{equation}
In the following discussion, we consider a non-degenerate system Hamiltonian. This assumption simplifies the mathematical derivations and provides a clearer illustration of the role of catalysts. In practice, however, physical Hamiltonians are often degenerate. Biswas reveals that degeneracies expand the permutation space within energy subspaces, thereby broadening the set of admissible thermal operations~\cite{biswas2023}. Scharlau et al find that for quantum systems with a fully degenerate Hamiltonian, all state transitions are exactly achievable using a bath no larger than the system~\cite{scharlau2018}. Koukoulekidis et al reveals that degeneracy permits the existence of multiple not fully passive states, in which the internal imbalance can be exploited to extract work~\cite{koukoulekidis2021}. Whether such degeneracies can further enhance the energy performance of quantum catalytic devices is an intriguing question worthy of future investigation.

\section{Optimal performances of a noncatalytic two-stroke refrigerator}
\label{III}
\begin{figure}		\includegraphics[width=0.35\textwidth]{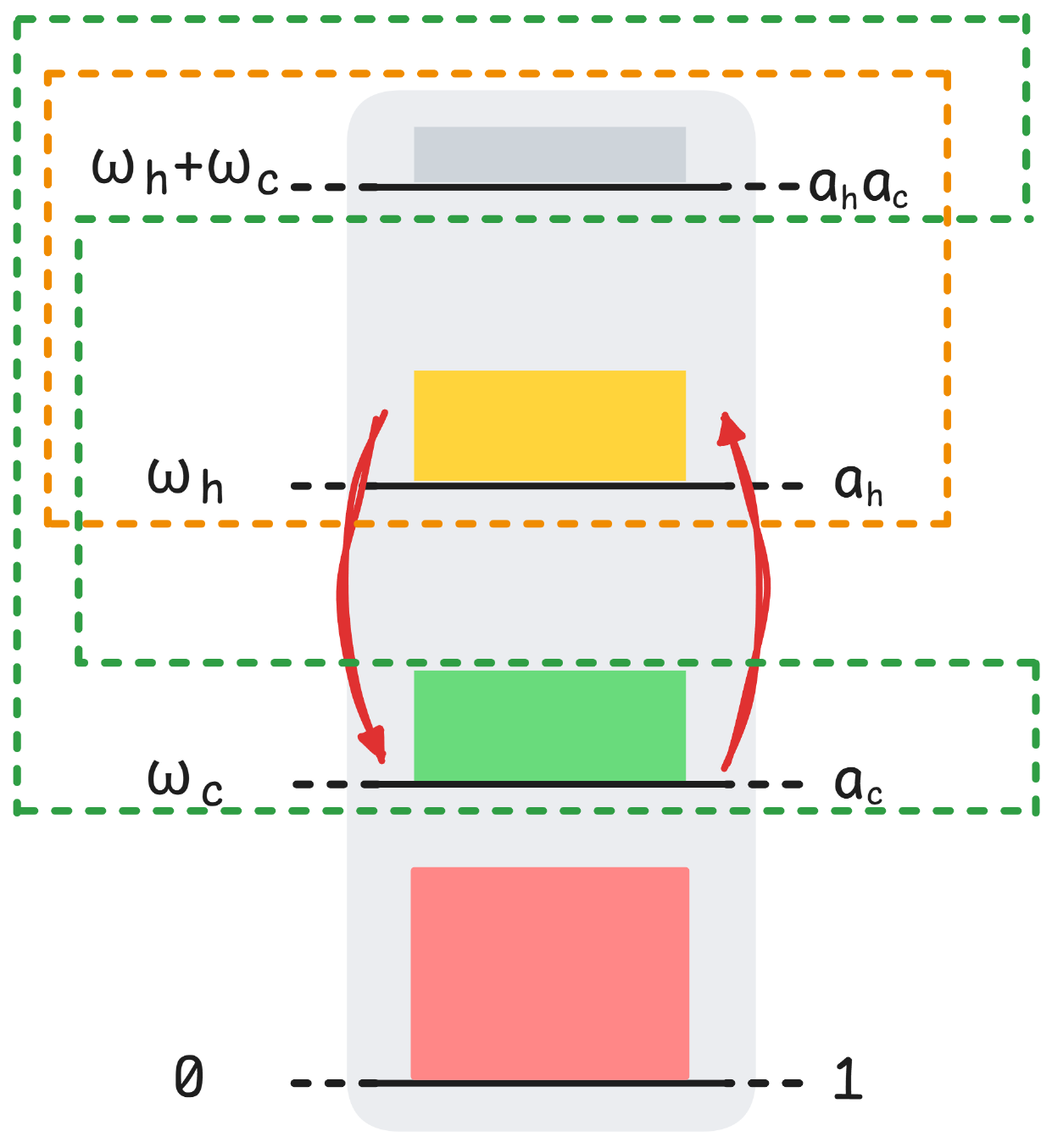}
    \caption{The schematic illustrates a permutation that exchanges the populations between the second and third excited states, resulting in an optimal coefficient of performance (COP) for the noncatalytic two-stroke refrigerator. }
    \label{fig:permutation_noncatalyst}
    \end{figure}

In this section, we calculate the optimal COP of a non-catalytic two-stroke refrigerator.  Identifying the optimal configuration is a highly complex and challenging task. Fortunately, the majorization theory applied in Ref.~\cite{lobejko2024,biswas2024} provides a means to simplify the computation of this process. In Appendix~\ref{appendix A}, we show that introducing certain permutation operations into the unitary evolution is sufficient to optimize both the maximum cooling power and the COP of the refrigerator within the majorization-based framework. This result thereby greatly simplifies the calculation of the optimal COP for a two-stroke refrigerator whose working medium consists of two TLSs. Consider a two-stroke refrigerator in which the initial state of the working body is given by $\tau_h \otimes \tau_c$. Given that the working medium has a dimension of $4$, the COP can be directly calculated for each of the $4! = 24$ possible permutations (see the Table \ref{tab:table1} in Appendix~\ref{appendix B}). Among all $24$ permutations, only four yield a positive quantity of heat absorbed from the cold reservoir, as listed in Table \ref{tab:table2} below.
\begin{figure*}
\includegraphics[width=0.8\textwidth]{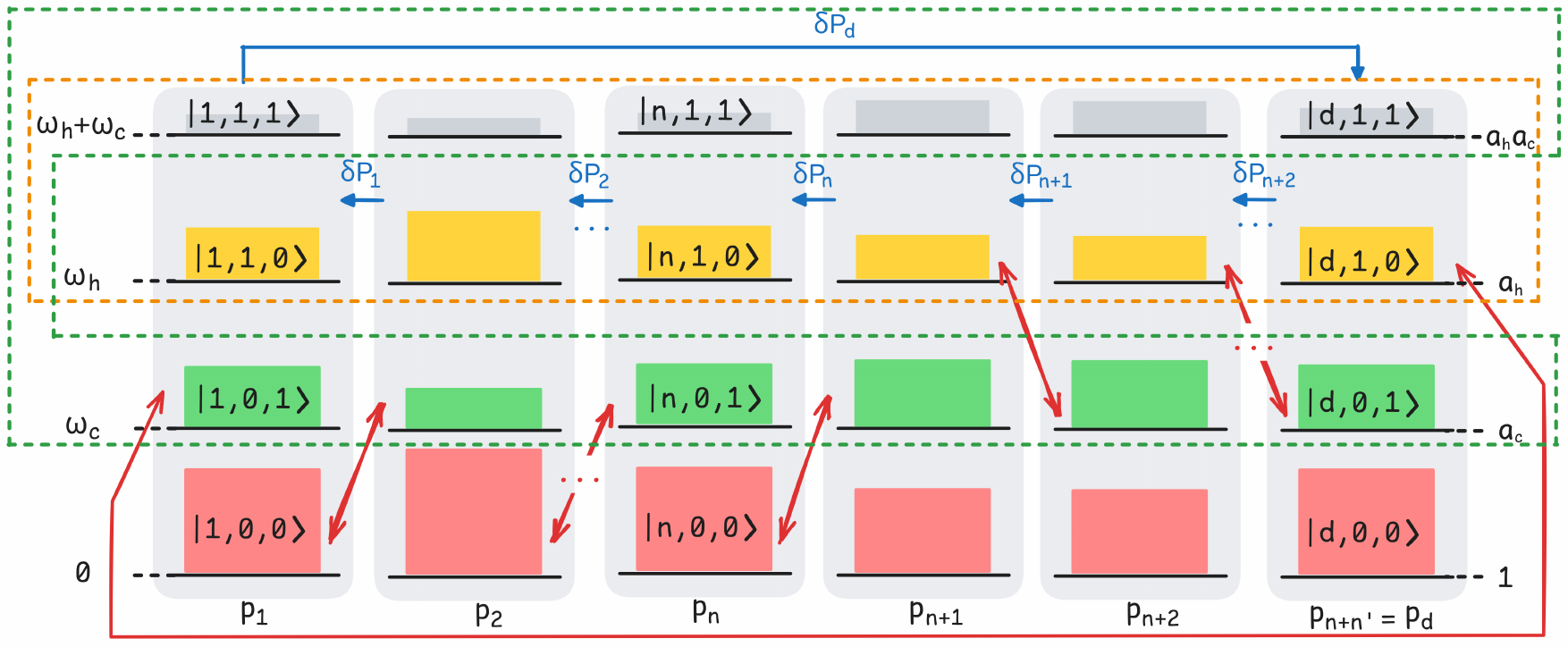}
    \caption{The permutation scheme employs a catalyst to enhance the COP of the refrigerator. The figure illustrates all energy levels of the composite system, where $|i,j,k\rangle \equiv |i\rangle_s |j\rangle_h |k\rangle_c$. The catalyst expands the original four energy levels (two TLSs) into $d$-node groups, where $d$ represents the dimension of the catalyst. Each column corresponds to the $i$ th catalyst state ($i \in [1,d]$) acts on original TLSs). The permutation of energy levels is illustrated with red arrows, which indicate the exchange of populations between the corresponding levels. The initial populations corresponding to the energy levels $|i,0,0\rangle_{s,h,c}$, $|i,0,1\rangle_{s,h,c}$, $|i,1,0\rangle_{s,h,c}$, $|i,1,1\rangle_{s,h,c}$ are $\frac{p_i}{(1+a_c)(1+a_h)}$, $\frac{p_ia_c}{(1+a_c)(1+a_h)}$, $\frac{p_ia_h}{(1+a_c)(1+a_h)}$, $\frac{p_ia_ca_h}{(1+a_c)(1+a_h)}$, respectively. The permutation between energy levels in the $i$ and $i+1$ node-groups lead to a net population flow form the $i$th node to the $(i + 1)$th, denoted by $\delta P_i$ and indicated with blue arrows. The region enclosed by the yellow dashed rectangle represents the hot subspace. Summing the populations within this area yields the total population of the excited state of the hot qubit. Similarly, the region enclosed by the green dashed line represents the cold subspace, and the sum of the populations within it corresponds to the excited-state population of the cold qubit.}
	\label{fig:permutation_1}
\end{figure*}
\begin{table*}
\caption{\label{tab:table2}
Four cases yield positive heat absorption from the cold reservoir in the noncatalytic two-stroke refrigerator}
\begin{ruledtabular}
\begin{tabular}{cccc}
Permutation matrices&
$Q_c$&
COP\\
\colrule
$|00\rangle\langle 00|+|01\rangle\langle 10|+|10\rangle\langle 01|+|11\rangle\langle 11|$ & $\frac{(a_c-a_h)\omega_c}{(1+a_h)(1+a_c)}$ & $\frac{\omega_c}{\omega_h-\omega_c}$\\
$|00\rangle\langle 00|+|01\rangle\langle 11|+|10\rangle\langle 01|+|11\rangle\langle 10|$ & $\frac{(a_c-a_h)\omega_c}{(1+a_h)(1+a_c)}$ & $\frac{\omega_c}{\frac{1-a_h}{1-a_h/a_c}\omega_h-\omega_c}$\\
$|00\rangle\langle 01|+|01\rangle\langle 10|+|10\rangle\langle 00|+|11\rangle\langle 11|$ & $\frac{(a_c-a_h)\omega_c}{(1+a_h)(1+a_c)}$ & $\frac{\omega_c}{\frac{1-a_h}{a_c-a_h}\omega_h-\omega_c}$\\
$|00\rangle\langle 01|+|01\rangle\langle 11|+|10\rangle\langle 00|+|11\rangle\langle 10|$ & $\frac{(a_c-a_h)\omega_c}{(1+a_h)(1+a_c)}$ & $\frac{\omega_c}{\frac{1-a_ca_h}{a_c-a_h}\omega_h-\omega_c}$\\
\end{tabular}
\end{ruledtabular}
\end{table*}

Therefore, the thermal machine operates as a refrigerator only in these four cases. We can then compare the COP corresponding to each of these four permutations and identify the optimal COP  with the aid of permutation (see Fig.~\ref{fig:permutation_noncatalyst})
\begin{equation}
\Pi_{\mathrm{opt}}=(|00\rangle\langle 00|+|01\rangle\langle 10|+|10\rangle\langle 01|+|11\rangle\langle 11|)_{h,c},
    \label{optimal_COP_permutation}
\end{equation}
where the first and second index in  $|\cdot\rangle\langle \cdot|$ correspond to the energy levels number of the hot and cold TLS of the working medium, respectively. According to Eqs.~\eqref{Qc} and \eqref{W}, the heat absorbed and the work consumed by the refrigerator when the working medium undergoes the optimal permutation transformation $\Pi_{\mathrm{opt}}$ can be denoted as follows:
\begin{equation}
Q_c=\mathcal{N} (a_c-a_h)\omega_c,
    \label{optimal_Qc_permutation}
\end{equation}
\begin{equation}
W=\mathcal{N}(\omega_h-\omega_c)(a_c-a_h),
    \label{optimal_W_permutation}
\end{equation}
where
\begin{equation}
a_{h/c} \equiv e^{-\beta_{h/c} \omega_{h/c}}, ~\mathcal{N} \equiv \frac{1}{(1+a_h)(1+a_c)}.
    \label{simplified_definition}
\end{equation}

Based on the above results, among the 24 possible energy permutation schemes, the optimal COP under non-catalytic conditions is given by 
\begin{equation}
\mathrm{COP}=\frac{\omega_c}{\omega_h-\omega_c},
    \label{optimal_COP}
\end{equation}
which is consistent with the COP of the \textit{Otto} cycle ~\cite{kosloff2014,Cangemi2024}.
In addition, for the refrigerator configuration, the conditions $W > 0$, $Q_c\ge 0$, and COP$\ge0$ must be satisfied, leading to:
\begin{equation}
	\beta_h \omega_h \geq \beta_c \omega_c.
	\label{eq:refri_condition}
\end{equation}

\section{Enhancement of the optimal COP via catalyst in two-stroke refrigerators}
\label{IV}
We have demonstrated that a specific unitary transformation among energy levels, defined as a permutation, can induce directed energy transfer from the cold to the hot reservoir. In this section, we show that the introduction of a catalyst further expands the set of accessible permutation pathways and enhances the optimal COP. 

As shown in the caption of Fig.~\ref{fig:permutation_1}, each energy level is associated with a specific population. Consequently, the permutation between the energy levels in the $i$ and the $(i+1)$ nodes lead to a net population flow from node $i$ to node $(i + 1)$, denoted by $\delta P_i$. For example, in the permutation between states $|1,0,0\rangle$ and  $|2,0,1\rangle$, a population of  $\mathcal{N}p_1$ flows from node $1$ to node $2$, while a population of   $\mathcal{N}p_2a_c$ flows from node $2$ to node $1$. Accordingly, a net population transfer 
\begin{equation}
\delta P_1=\mathcal{N}(p_2a_c-p_1)
    \label{deltaP_1}
\end{equation}
occurs from node $1$ to $2$. Crucially, to ensure that the marginal state of the catalyst remains unaffected by the permutation performed during the work stroke, the net population entering a given node must equal the population leaving it, thereby restoring the catalyst to its original state. Through analysis of the population flows corresponding to the remaining nodes, we arrive at the following results $\delta P_1=\delta P_2= \cdots \delta P_d=\delta P$. To quantify the heat current generated by the net population flow, we define the hot and cold subspaces~\cite{lobejko2024}
\begin{equation}
\mathcal{H}=\mathrm{set}\{ \forall i,k;j=1\quad |i,j,k\rangle_{s,h,c} \}  \label{hot_subspace}
\end{equation}
and
\begin{equation}
\mathcal{C}=\mathrm{set}\{ \forall i,j;k=1\quad |i,j,k\rangle_{s,h,c} \},
\label{cold_subspace}
\end{equation}
where $i,~j,~k$ represent the energy levels of the Hamiltonian of the composite working body $H_s+H_h+H_c$. For example, Fig.~\ref{fig:permutation_1} illustrates the excited hot and cold subspaces of a two-stroke refrigerator, enclosed by a yellow dashed rectangle and a green dashed line, respectively. Within the framework where the ground-state energy is set to zero and considering the partitioning of subspaces, the heat $Q_c$ [according to Eq.~\eqref{Qc}] absorbed from the cold reservoir can be rigorously expressed as the sum of net population outflows from the excited cold subspace, each weighted by the corresponding excitation energy  $\omega_c$. Mathematically, this is given by 
\begin{equation}
Q_c=\sum_m \delta P_m \omega_c,
\label{cold_subspace_Qc}
\end{equation}
where the summation index $m$ runs over all permutations within the cold subspace. Analogously, the heat $Q_h$ [Eq.~\eqref{Qh}] dissipated to the hot reservoir is obtained by summing the net population inflows into the excited hot subspace, each weighted by the corresponding energy gap $\omega_h$ of the excited level, i.e.,
\begin{equation}
Q_h=\sum_{m^\prime} \delta P_{m^\prime} \omega_h,
\label{hot_subspace_Qh}
\end{equation}
where the summation index  $m^{\prime}$ runs over all permutations within the hot subspace. 

To enhance the COP of the refrigerator after introducing the catalyst, it is essential to maximize heat absorption while minimizing heat release. Accordingly, we consider the permutation operation illustrated in Fig.~\ref{fig:permutation_1} , which can be mathematically represented as follows: 
\begin{align}
   \Pi_1&=(\sum_{m=1}^{n}|m,0,0\rangle\langle m+1,0,1|+\sum_{m^{\prime}=n+1}^{n+n^{\prime}-1}|m^{\prime},1,0\rangle\langle m^{\prime}+1,0,1|\nonumber\\
   &+ |n+n^{\prime},1,0\rangle\langle 1,0,1|+\text { Herm. conjugate } )_{s,h,c}+\tilde{\mathbb{I}}_{\mathrm{Rest}},
   \label{permutation_1}
\end{align}
\begin{figure*}
\includegraphics[width=0.8\textwidth]{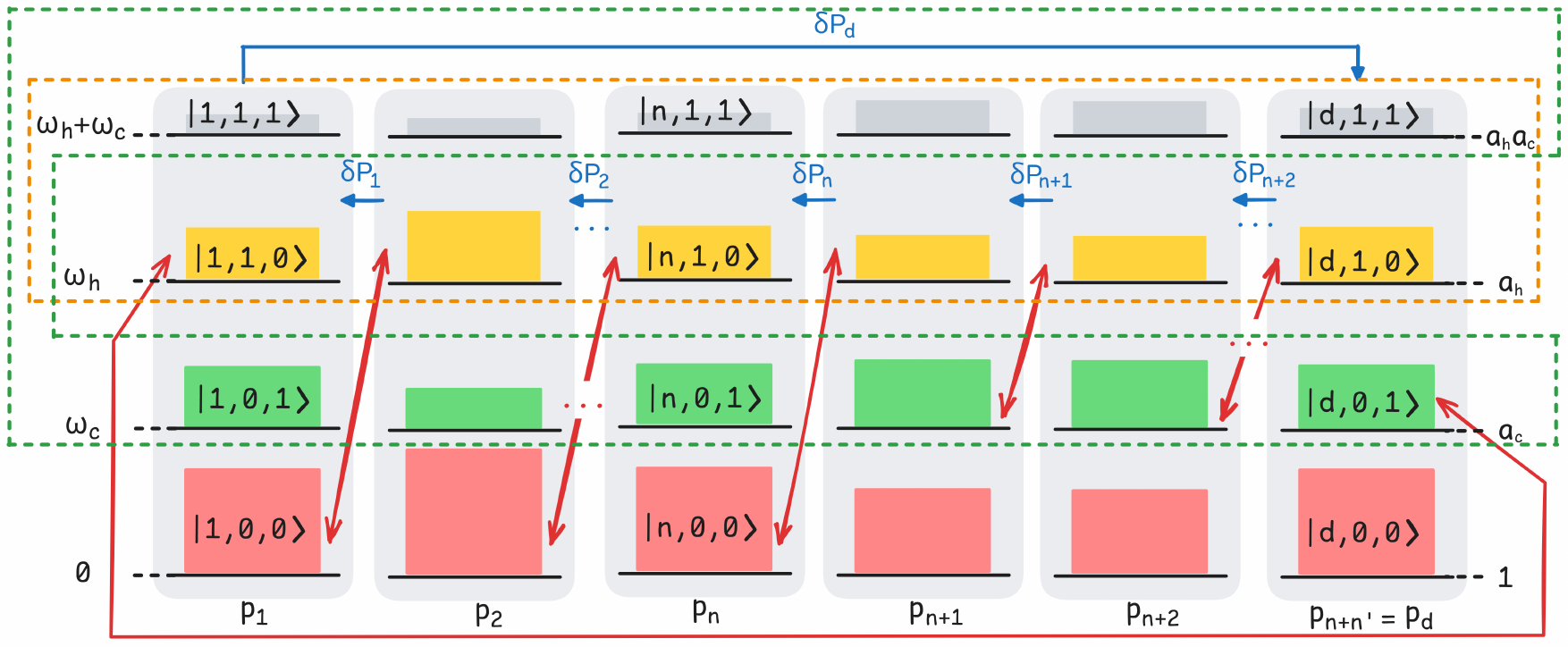}
    \caption{The energy level permutation scheme that utilizes a catalyst to expand the operation regime of the refrigerator.}
	\label{fig:permutation_2}
\end{figure*}
where $d=n+n^{\prime}$ denotes the dimension of the catalyst, $\mathbb{I}_{\mathrm{Rest}}$ is the identity operator acting on the subspace orthogonal to the first term in the parentheses, and the Hermitian conjugate term ensures the completeness of the permutation. $n$ denote the number of catalyst levels participating in the permutations between the cold subspace and the ground states (acting between $|0,0\rangle$ and $|0,1\rangle$). The remaining $n^{\prime}=d-n$ catalyst levels are associated with the permutations between the hot and cold subspaces (acting between $|1,0\rangle$ and $|0,1\rangle$). $n^{\prime}$ is a positive integer whose maximum value does not exceed the catalyst dimension. To calculate the heat, we assume that the permutation $\Pi_1$ induces a net population flow of $\delta P$ from a node to its left. Consequently, there is a net outflow of $d \delta P$ from the cold subspace, while a net inflow of $n^{\prime} \delta P$ occurs into the hot subspace. With Eqs.~\eqref{COP},~\eqref{cold_subspace_Qc} and \eqref{hot_subspace_Qh}, we have 
\begin{equation}
\mathrm{COP}=\frac{d \omega_c \delta P}{ n^{\prime}\omega_h \delta P-d \omega_c \delta P}=\frac{\omega_c}{ \frac{n^{\prime} \omega_h}{d}-\omega_c}.
\label{subspace_COP}
\end{equation}

\begin{figure*}
\includegraphics[width=0.95\textwidth]{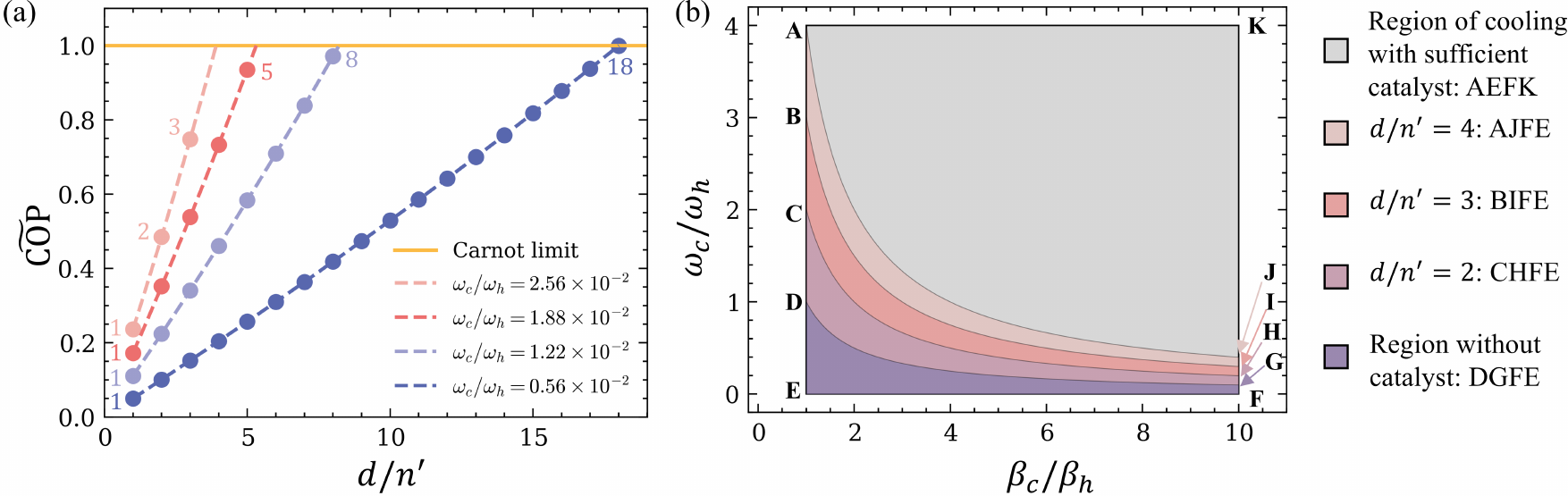}
    \caption{(a) Normalized coefficient of performance $\tilde{\mathrm{COP}} \equiv \mathrm{COP}/\mathrm{COP}_{\mathrm{Carnot}}$ versus $d/n^{\prime}$, where  $\mathrm{COP}_{\mathrm{Carnot}}= \beta_h/(\beta_c-\beta_h) $ and $\beta_c/\beta_h=10$. The color gradient from dark blue to pink represents increasing energy level ratio $\omega_c/\omega_h = 0.56,\ 1.22,\ 1.88,\ 2.56\ (\times 10^{-2})$. The Carnot COP is shown as a yellow solid line. The dotted line indicates the upward trend of COP. Integer points along the dotted line represent actual achievable efficiencies, with the corresponding catalyst dimensions labeled accordingly. (b) The operating region of cooling, determined by the energy level ratio and temperature ratio, lies beneath the boundary (e.g., the region CFHE). Distinct colors represent the operating ranges associated with various $d/n^\prime$.}
	\label{fig:catalyst_enhancement}
\end{figure*}

\begin{figure}
\includegraphics[width=0.5\textwidth]{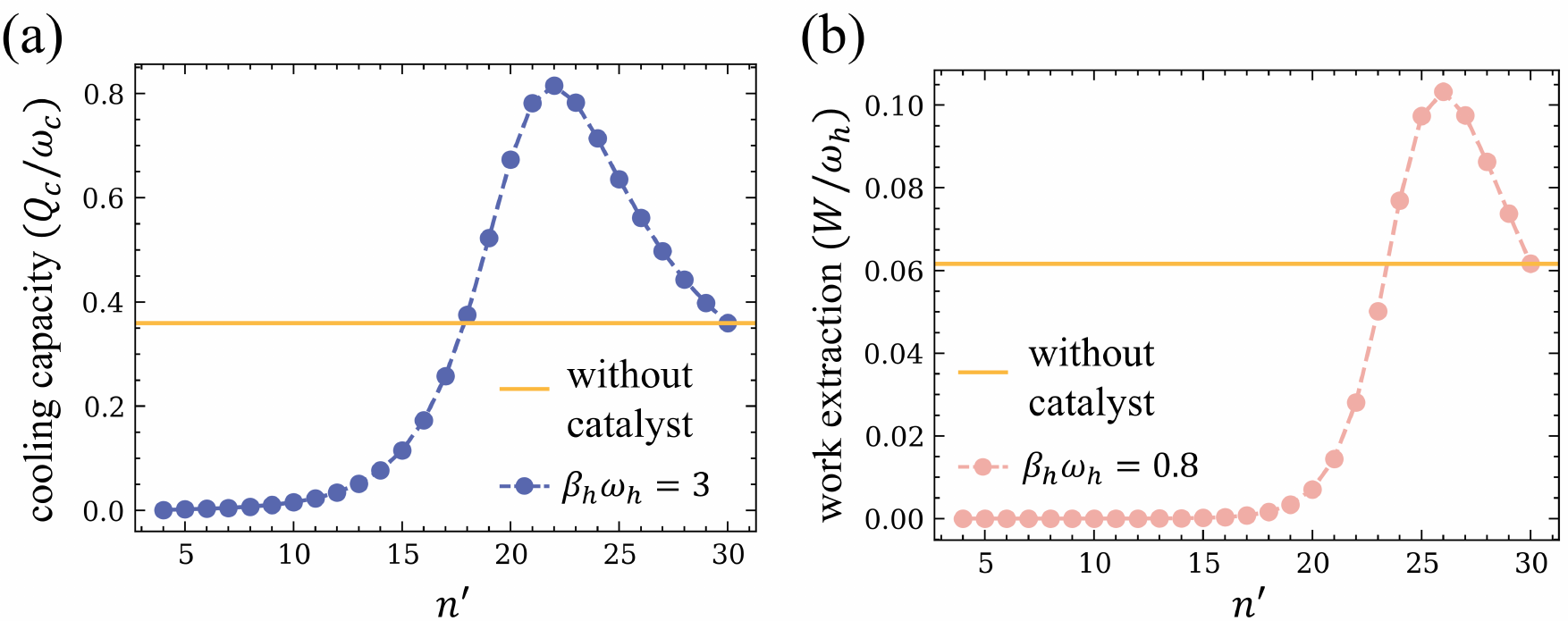}
    \caption{(a) This figure depicts the variation of the cooling capacity with $n^{\prime}$ as formulated in Eq.~\eqref{cooling capacity} with $\frac{\beta_h \omega_h}{\beta_c \omega_c}=8$. Here, we have taken the  dimension of catalyst is $30$. (b) The variation of the work produced by the catalyst-assisted two-stroke engine with $n^{\prime}$ as formulated in Ref.~\cite{biswas2024}.}
	\label{fig:comparison}
\end{figure}

In practice, for $n^\prime=1$, adjusting the parameter $d$ is the most effective way to enhance the COP, albeit only yielding discrete COP values. Introducing $n^\prime$ substitutions not only improves the tunability of the heat engine during operation, but also allows the refrigeration capacity to be regulated by varying $n^\prime$, leading to an optimized trade-off, as shown in Fig.~\ref{fig:comparison}. The choice of $n^{\prime}/d$ is governed by the second law of thermodynamics. In the context of thermodynamics, entropy production plays a crucial role in understanding the efficiency limits of quantum machines. After completing a closed cycle, the working medium returns to its initial state without a change in entropy. However, due to the inherent irreversibility of the process, the entropy production $\sigma$ must satisfy the inequality:
\begin{align}
\sigma=-\beta_{c} Q_{c}+\beta_{h} Q_{h}\geq 0.
   \label{2nd_prove_2}
\end{align}
\begin{figure}
\includegraphics[width=0.4\textwidth]{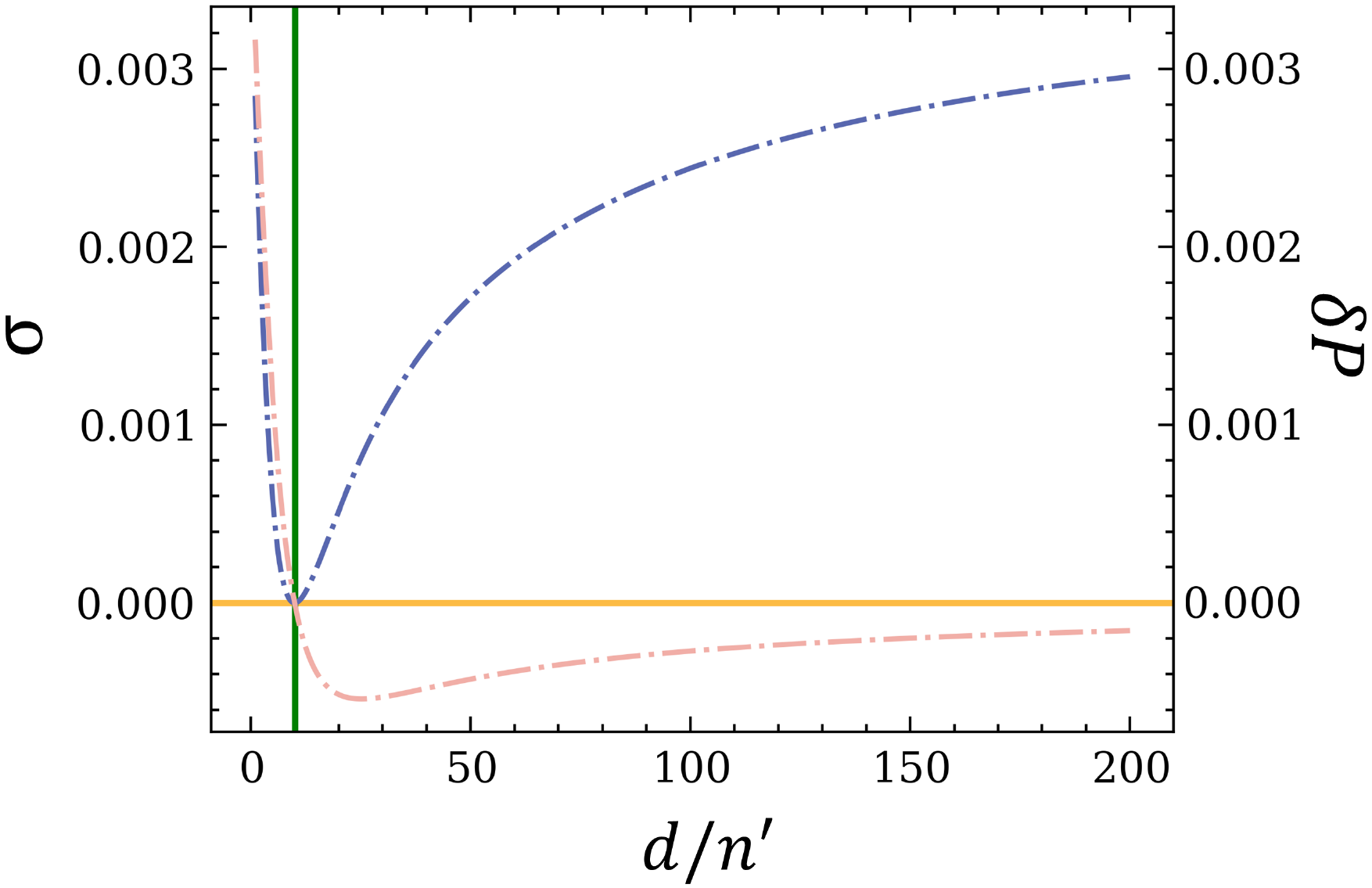}
    \caption{The entropy production $\sigma$ (the left axis) and  $\delta P$ (the right axis) varying with the catalyst dimension $d/n'$. The blue and pink dash-dotted curve plots $\sigma$ from Eq.~\eqref{2nd_prove_2} and $\delta P$ from Eq.~\eqref{delta_P}, respectively. The vertical green solid line indicates the theoretical critical point $d/n^{\prime} = \beta_h \omega_h / (\beta_c \omega_c)$, and the horizontal yellow solid line marks the condition $\sigma=0$.}
	\label{fig:2nd}
\end{figure}
Furthermore, we verify the thermodynamic consistency of our model through numerical calculations, as shown in Fig.~\ref{fig:2nd}. The non-negative entropy production confirms that our model is in agreement with the second law, ensuring thermodynamic consistency. In Fig.~\ref{fig:2nd}, the vertical green line divides the diagram into two regions: the refrigerator regime on the left (where $\delta P > 0$) and the heat-engine regime on the right (where $\delta P < 0$). A detailed classification of the efficiency is provided in the Appendix~\ref{appendix A}. The intersection of the yellow horizontal line ($\sigma=0$) with the green vertical line coincides with the critical point at which both $\sigma$ and $\delta P$ vanish. For the permutation $\Pi_1$ and within the operational range of the refrigerator ($\delta P >0$), we have $Q_c=d \omega_c \delta P$ and $Q_h=n^{\prime} \omega_h \delta P$. After simplification, we obtain an inequality 
\begin{equation}
d/n^{\prime} \le \frac{\beta_h \omega_h}{\beta_c \omega_c},
\label{d_condition_1}
\end{equation}
which also guarantees that the COP does not exceed the Carnot limit $\mathrm{COP}_{\mathrm{C}}\equiv\beta_h/(\beta_c-\beta_h)$. As illustrated in Fig.~\ref{fig:catalyst_enhancement}(a), when $d=n^{\prime}=1$, it corresponds to the optimal Otto cycle COP in the non-catalytic case. As $d/n^{\prime}$ increases, the COP rises monotonically, owing to the enhanced heat flux resulting from the contribution of additional cold subspaces. The maximum dimension is bounded by the Carnot COP. When the temperature ratio $\beta_c/\beta_h$ and the catalyst dimension $d$ are fixed, an increase in the energy level ratio  $\omega_h/\omega_c$ leads to a reduced probability of high-temperature excited states. This, in turn, frees up more excited-state space for low-temperature population, thereby enhancing the COP. 

We make a remark here: Analogous unitary operations have been realized in various quantum platforms via engineered Hamiltonians~\cite{abah2012,camati2016,masuyama2018,barker2022,zhou2023}. Specifically, longitudinal driving enables full and coherent energy-level manipulation in multilevel quantum systems, including effective permutation of level structures via synthetic couplings~\cite{zhou2023}. In contrast, self-contained thermal machines provide a feasible alternative by operating autonomously without external driving, as demonstrated within devices such as cavity quantum electrodynamics systems~\cite{junior2024} and superconducting circuits~\cite{biswas2023,linden2010,yu2019,aamir2025a}.

\begin{figure}
\includegraphics[width=0.35\textwidth]{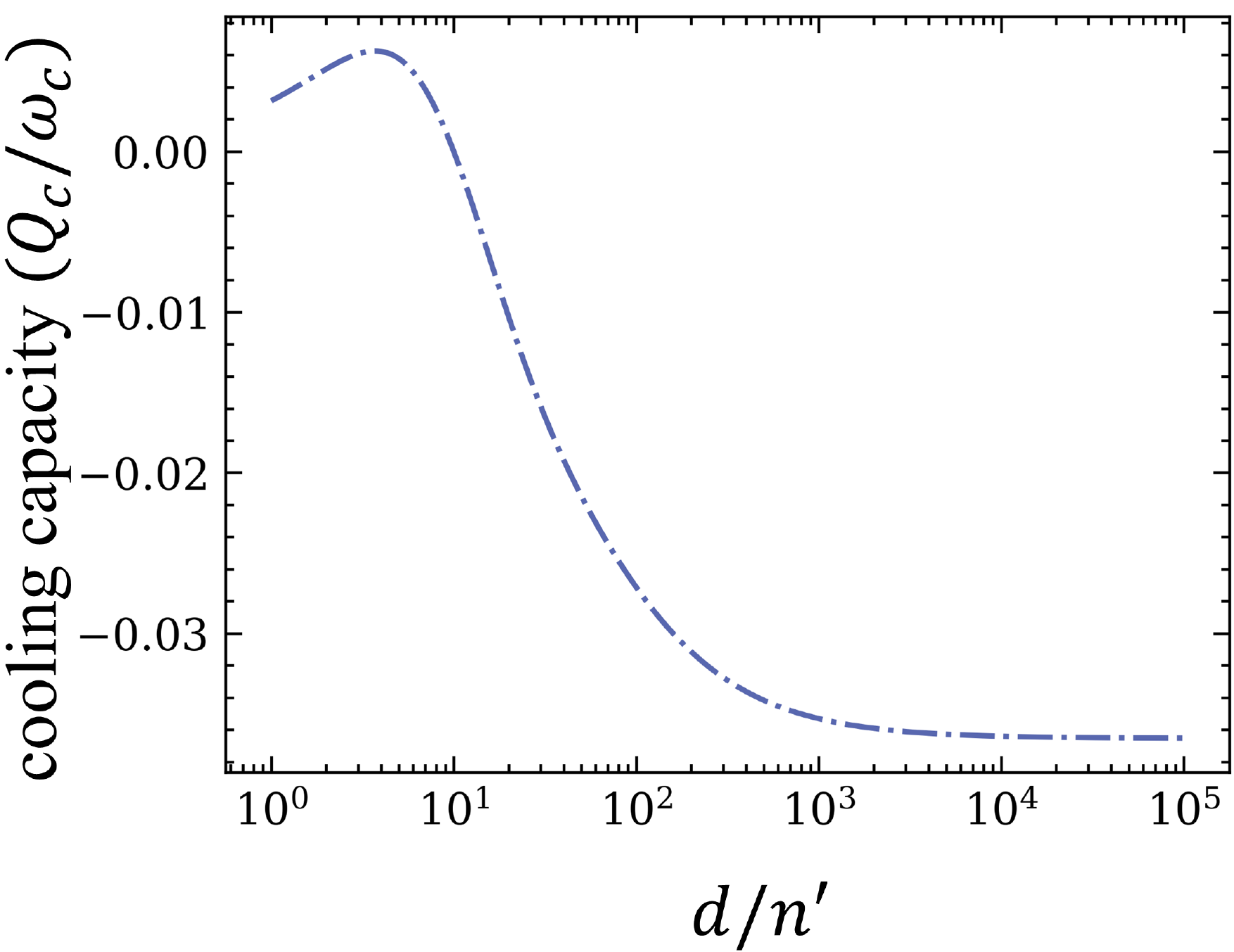}
    \caption{The dimensionless cooling capacity $Q_c/\omega_c$ as a function of the catalyst dimension $d/n^{\prime}$. The horizontal axis is plotted on a logarithmic scale. In the refrigerator regime ($\delta P > 0$), $Q_c$ increases and reaches a maximum, after which it decreases and eventually saturates to a finite value in the heat-engine regime ($\delta P < 0$) as $d/n^{\prime}$ continues to grow.}
	\label{fig:3rd}
\end{figure}

To determine the heat absorption during the work stroke mediated by the permutation operation $\Pi_1$ given in Eq.~\eqref{permutation_1}, we need systematically derive $\delta P$ as a function of the reservoir inverse temperatures $\beta_h$ and $\beta_c$, the energy gaps $\omega_h$ and $\omega_c$ characterizing the hot and cold two-level systems, and the dimension of the catalyst $d=n+n^{\prime}$. The derivation proceeds by formulating the catalyst preservation constraints, and the explicit calculations are presented in Appendix~\ref{appendix C}. Notably, we obtain
\begin{equation}
\delta P=\frac{1}{f(a_c,a_h,n,n^{\prime}) }\mathcal{N}(a_c^{n+n^{\prime}}-a_h^{n^{\prime}}).
\label{delta_P}
\end{equation}
Therefore, the constraints for catalyst preservation limits the unique value of $\delta P$ and $p_i$, which is reasonable because, once the configuration and temperature are determined, the COP and cooling capacity
\begin{equation}
Q_c=d \omega_c \delta P
\label{cooling capacity}
\end{equation}
is also determined. When the temperature of reservoirs $\beta_h$ and $\beta_c$ and the energy gaps $\omega_h$ and $\omega_c$ of the two TLSs are fixed, the cooling capacity $Q_c$ can be optimized by adjusting the ratio $n/n'$ to modify the initial probability distribution $p_i$ of the catalyst, which will be illustrated in Fig.~\ref{fig:comparison}(a). Furthermore, as shown in Fig.~\ref{fig:2nd}, $\delta P$ decreases with increasing $d$, passing through zero at the critical dimension where the device transitions from operating as a refrigerator ($\delta P > 0$) to functioning as a heat engine ($\delta P < 0$). Consequently, the cooling rate $Q_c$ cannot increase arbitrarily and vanishes at this transition point. It should be further noted that in the heat engine regime, $Q_c$ remains finite and ultimately saturates to a finite value as the catalyst dimension increases, as depicted in Fig.~\ref{fig:3rd}.

\section{Catalytic enhancement for extending the operational range of a two-stroke refrigerator}
\label{V}

The performance of two-stroke quantum refrigerators is typically constrained within a limited range of control parameters, such as energy level spacing, coupling strength, or reservoir temperatures. Beyond this range, the device may cease to operate effectively as a refrigerator. This limitation restricts both the practical applicability and the flexibility of such thermal machines. In the section, we will show that quantum catalysis offers a novel mechanism to overcome this constraint. By introducing the catalyst, it becomes possible to expand the operational parameter space of the refrigerator. 

In order to achieve a broader operational regime for the refrigerator, we consider an alternative permutation
\begin{align}
   \Pi_2&=(\sum_{m=1}^{n}|m,0,0\rangle\langle m+1,1,0|+\sum_{m^{\prime}=n+1}^{n+n^{\prime}-1}|m^{\prime},0,1\rangle\langle m^{\prime}+1,1,0|\nonumber\\
   &+ |n+n^{\prime},0,1\rangle\langle 1,1,0|+\text { Herm. conjugate } )_{s,h,c}+\tilde{\mathbb{I}}_{\mathrm{Rest}},
   \label{permutation_2}
\end{align}
as illustrated in Fig.~\ref{fig:permutation_2}. Note that $\tilde{\mathbb{I}}_{\mathrm{Rest}}$ denotes the identity operator on the orthogonal complement of the first term in the parentheses. Here, $n$ represents the number of catalyst levels participating in the permutations between the hot subspace and the ground states (i.e., between $|0,0\rangle$ and $|1,0\rangle$). The remaining $n^{\prime}=d-n$ catalyst levels are associated with permutations between the cold and hot subspaces (i.e., between $|0,1\rangle$ and $|1,0\rangle$). As mentioned earlier, in order to preserve the marginal state of the catalyst, all net population transfers must be balanced, assuming a fixed value of $\delta P^{\prime}$. Employing Eqs.~\eqref{COP},~\eqref{cold_subspace_Qc} and \eqref{hot_subspace_Qh}, we can calculate
\begin{equation}
\mathrm{COP}^{\prime}=\frac{n^{\prime} \omega_c \delta P^{\prime}}{(n+n^{\prime})\omega_h \delta P^{\prime}-n^{\prime} \omega_c \delta P^{\prime}}=\frac{\omega_c}{\frac{d\omega_h}{n^{\prime}}-\omega_c}.
\label{subspace_COP_2}
\end{equation}
A direct consequence of the second law of thermodynamics, as expressed in Eq.~\eqref{2nd_prove_2} for permutation $\Pi_1$, also results in
\begin{equation}
0 \le \mathrm{COP}^{\prime}\le \frac{\beta_h}{\beta_c-\beta_h}.
\label{refri_cond}
\end{equation}
By straightforwardly substituting Eq.~\eqref{subspace_COP_2} into Eq.~\eqref{refri_cond}, we find
\begin{equation}
\frac{d}{n^{\prime}} \ge \frac{\omega_c \beta_c}{\omega_h \beta_h}.
\label{refri_cond_simplified}
\end{equation}
Under this constraint, the cooling operating region for various $d/n^{\prime}$ is illustrated in Fig.~\ref{fig:catalyst_enhancement}(b). When there is no catalyst, $d/n^{\prime} = 1$ and the region of cooling is given by the area DGFE. With the aid of a catalyst and adjustment of the heat flux ratio between the hot and cold subspaces, the magnitude of $d/n^{\prime}$ can be regulated. An increase in $d$ shifts the operational cooling boundary outward, thereby expanding the enclosed operational region. As $d/n^{\prime}$ tends to infinity, the operational range approaches the rectangular region denoted by AEFK, since the sole requirement for refrigeration in this limit is $\beta_c > \beta_h$. From the inequality Eq.~\eqref{refri_cond_simplified}, we can tune $d/n^{\prime}$ to ensure the operation of refrigerator even if $\omega_c > \omega_h$.

\section{Performance Comparison of Catalytically Enhanced Refrigerators and Heat Engines}
\label{VI}

Inspired by previous works on catalytic enhancement of heat engine performance~\cite{lobejko2024,biswas2024}, we further explore its application to refrigeration systems. Due to the differences in the direction of the thermodynamic cycle and system functionality, the catalytic protocols and resulting performance characteristics in refrigerators differ from those in heat engines. In this section, we quantitatively compare and elucidate the origins of these distinctions.

The first difference lies in the catalytic strategy: whereas a single permutation in the heat engine achieves simultaneous improvement in both efficiency and working range, the refrigerator requires two distinct permutations to attain comparable performance enhancement.

This difference can be demonstrated through mathematical proof. First, according to Ref.~\cite{lobejko2024,biswas2024}, the operating condition of heat engine without catalyst is given as follows
\begin{equation}
\beta_h\omega_h \le \beta_c\omega_c~~\text{and}~~\omega_h>\omega_c.
    \label{the operating condition of heat engine}
\end{equation}
When the catalyst is introduced, its efficiency can be expressed in the following form $\eta=1-h(d)\frac{\omega_c}{\omega_h}$, where $h(d)$ is a function of $d$. Further derivation shows that enhancing the efficiency requires satisfying condition 
\begin{equation}
1-\frac{\omega_c}{\omega_h}<\eta<1-\frac{\beta_h}{\beta_c} \Rightarrow \frac{\beta_h\omega_h}{\beta_c\omega_c} < h(d) < 1.
    \label{enhancing the efficiency}
\end{equation}
By introducing a $d$-dimension catalyst and utilizing the expression for its efficiency $\eta$, the operation regime can be determined as 
\begin{equation}
0 < 1-h(d)\frac{\omega_c}{\omega_h} <1-\frac{\beta_h}{\beta_c},
    \label{operating range eta}
\end{equation}
which can be further simplified as
\begin{equation}
\frac{\beta_c\omega_c}{\beta_h\omega_h} > \frac{1}{h(d)}~~\text{and}~~\frac{\omega_c}{\omega_h} < \frac{1}{h(d)}.
    \label{expanding the operating range}
\end{equation}
From Eq.~\eqref{enhancing the efficiency}, we can obtain that the condition for enhancing the efficiency extends the boundary of Eq.~\eqref{expanding the operating range}, specifically $1/h(d)$, beyond a value of $1$. From a physical illustration perspective, as shown in Fig.~\ref{fig:catalyst_enhancement}(b), with $\beta_c/\beta_h$ and $\omega_c/\omega_h$ representing the horizontal and vertical axes, respectively, the first operational constraint in Eq.~\eqref{the operating condition of heat engine} requires that the heat engine can only operate above the hyperbola $\frac{\beta_c\omega_c}{\beta_h\omega_h}=1$, while the second confines it below the horizontal line $\frac{\omega_c}{\omega_h}=1$. Thus, the actual operational region is confined between these two boundaries. The condition for enhancing efficiency, $h(d)<1$, implies that the effective operational boundary is extended, thereby broadening the accessible region. Thus, a single substitution operation can simultaneously enhance both the efficiency and the operating range of the heat engine.

We next consider the case of the refrigerator without catalyst, whose operating condition is given below
\begin{equation}
\beta_h\omega_h \ge \beta_c\omega_c~~\text{and}~~\omega_h>\omega_c.
    \label{the operating condition of refri}
\end{equation}
With the catalyst, the COP takes the following form $\mathrm{COP}=\frac{\omega_c}{h^\prime (d)\omega_h-\omega_c}$, where $h^\prime(d)$ is a function of $d$. Here, improving the COP requires condition
\begin{equation}
\frac{\omega_c}{\omega_h-\omega_c}<\mathrm{COP}<\frac{\beta_h}{\beta_c-\beta_h} \Rightarrow \frac{\beta_c\omega_c}{\beta_h\omega_h} < h^\prime(d) < 1,
    \label{enhancing the COP}
\end{equation}
By introducing a $d$-dimension catalyst and utilizing the expression for its COP, the operation regime can be determined as 
\begin{equation}
0 < \frac{\omega_c}{h^\prime (d)\omega_h-\omega_c} <\frac{\beta_h}{\beta_c-\beta_h},
    \label{operating range COP}
\end{equation}
which can be further simplified as
\begin{equation}
\frac{\beta_c\omega_c}{\beta_h\omega_h} < h^\prime(d),
    \label{enhancing the range}
\end{equation}
where $h^\prime(d)<1$ for the condition of enhancing COP. Similarly, from the perspective of the physical diagram, the first constraint restricts the refrigerator to operate only below the hyperbola $\frac{\beta_c\omega_c}{\beta_h\omega_h}=1$, while the second confines it beneath the horizontal line $\frac{\omega_c}{\omega_h}=1$. Since the first constraint is more restrictive, the actual accessible region lies below the curve $\frac{\beta_c\omega_c}{\beta_h\omega_h}=1$. The condition for enhancing the COP, which requires $h^\prime(d) < 1$, implies that the boundary constraint is tightened as described by Eq.~\eqref{enhancing the range}, leading to a reduction in the operational range. Therefore, it is shown that these two conditions are mutually exclusive. Hence, no single substitution can simultaneously improve both the COP and the operating range. 

Fortunately, as in the heat engine case, the refrigerator also exhibits a simultaneous enhancement in both cooling capacity and COP under permutation $\Pi_1$, as demonstrated in Fig.~\ref{fig:comparison}. This suggests that a more favorable trade-off between these quantities can be achieved by appropriate selection of $n^\prime$ and the catalyst dimension $d$. In Fig.~\ref{fig:comparison}(a), we plot $Q_c$ as a function of $n^\prime$ for a fixed total dimension $d = n + n^\prime = 30$. Notably, for certain values of $n^\prime$, the cooling heat extracted by the catalytic two-stroke refrigerator exceeds that of the non-catalytic version. Moreover, at rightmost point of the x-axis (i.e., at $n^\prime=30$) the COP of the catalyst-assisted two-stroke refrigerator matches with the Otto COP whereas for $n^\prime<30$ the COP is strictly greater than Otto COP. 

\section{Conclusions}
\label{VII}
This work demonstrates that the integration of a finite-dimensional catalyst into a two-stroke quantum refrigerator significantly enhances both its COP and cooling capacity, while also expanding its operational regime beyond conventional thermodynamic limits. Specifically, by leveraging catalytic majorization theory and permutation protocols, we show that the catalyst enables the refrigerator to exceed the Otto-bound COP and $Q_c$ and operate in previously inaccessible frequency and temperature ranges. The analysis reveals that two distinct permutation types are necessary to simultaneously improve COP and operational range in refrigerators—unlike heat engines, where a single permutation suffices. This highlights the unique role of catalysis in quantum thermal machines and provides a concrete framework for designing high-performance quantum cooling systems. Moreover, extending the performance limits of catalytic quantum refrigerators can serve to witness quantum properties~\cite{deoliveirajunior2025}.

Despite these promising results, several challenges and opportunities for future work remain. A key challenge is the incorporation of finite-time dynamics into the catalytic refrigerator model, which requires reconciling non-equilibrium thermodynamics with catalytic constraints under rapid cycling. This includes characterizing non-adiabatic dissipation, incomplete thermalization, and ensuring catalyst stability during fast operations. Additionally, the role of quantum coherence and correlations in catalytic processes warrants deeper investigation, potentially through quantum optimal control theory. Future research could also explore the effects of degenerate energy levels and higher-dimensional catalysts on performance limits. Experimental realizations in platforms such as superconducting circuits or trapped ions would be crucial for validating these theoretical advances and could lead to practical applications in quantum computing thermal management and high-efficiency microscale energy conversion. Ultimately, bridging these theoretical insights with experimental capabilities will be essential for harnessing catalytic quantum thermodynamics in real-world technologies.

\section*{Acknowledgment}

Cong Fu acknowledges Meiling Yan for insightful discussions and careful check. This work has been supported by the Natural Science Foundation of Fujian Province (2023J01006), National Natural Science Foundation of China (12364008 and 12365006), and Fundamental Research Fund for the Central Universities (20720240145).

\appendix

\section{Enhancing thermodynamic performances through of permutations in unitary operations}
\label{appendix A}
\begin{figure}[b]
\includegraphics[width=0.5\textwidth]{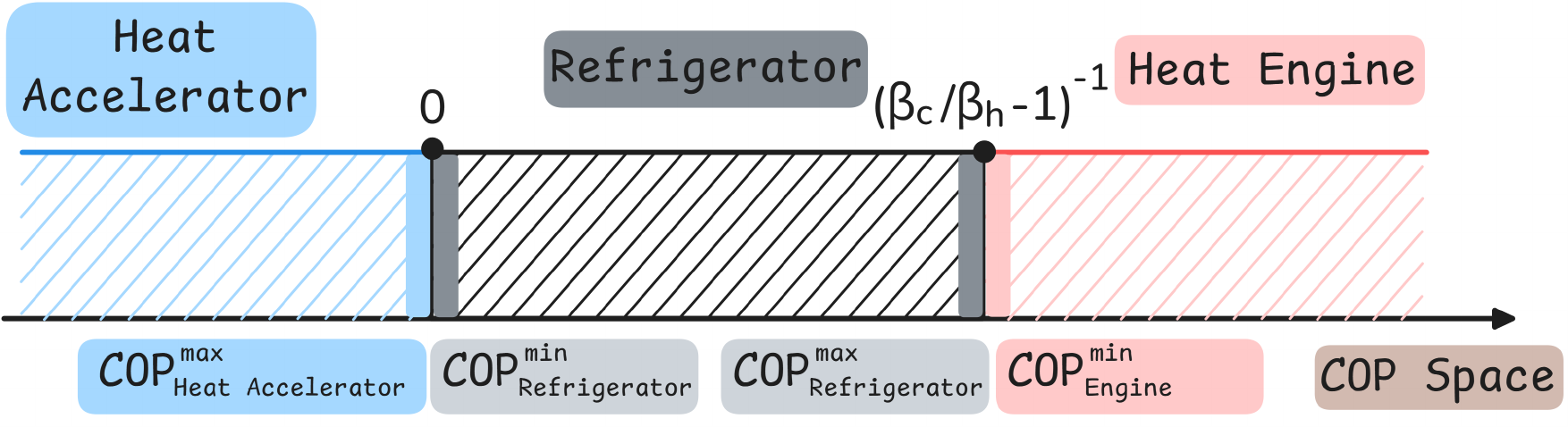}
    \caption{The ordering of the COP among the different
operational modes}
	\label{fig:COP_order}
\end{figure}

From Eq.~\eqref{Qc}, it is observed that $Q_c$ is a linear function of $\rho$ and does not depend on the off-diagonal terms of the state $\rho ^\prime$. Thus, the heat extracted from the cold reservoir can be 
rewritten as
\begin{equation}
Q_c=\mathrm{Tr}\left[H_c(\rho -\mathcal{D}(U\rho U^{\dag})) \right],
    \label{Qc_linear}
\end{equation}
where $\mathcal{D(.)}$ denotes the dephasing in the eigenbasis of the Hamiltonian $H_h+H_c+H_s$. Note that initial state $\rho$ is diagonal in the energy eigenbasis and the final state after a unitary evolution $U$ is given by $\rho^\prime = U\rho U^\dagger$. By Birkhoff's theorem~\cite{horn2012}, the eigenvalue vector of a Hermitian matrix majorizes its diagonal part. Therefore, we have the majorization relation as
\begin{equation}
|\rho^\prime\rangle \rangle \succ  |\mathcal{D}(\rho^\prime)\rangle \rangle.
    \label{Birkhoff1}
\end{equation}
Here, we give a definition on the notation $|\cdot\rangle \rangle$~\cite{sagawa2022}: Let $A$ be an operator with a spectral decomposition $A=\Sigma_m \lambda_m |m\rangle\langle m|$, where $\lambda_m$ are the eigenvalues of $A$ and $|m\rangle$ form an orthonormal eigenbasis of the Hilbert space. We denote the spectrum $|m\rangle$ by the bra-ket notation with double brackets, i.e., $|A\rangle \rangle$. We emphasize that this notation is used throughout this work to represent the spectrum of operator $A$. This differs from the more conventional usage in quantum information theory, where double brackets denote the vectorization of an operator in Liouville space under the Hilbert-Schmidt inner product. Our subsequent use of the inner product $\langle\langle A|B\rangle\rangle$ should thus be interpreted as a scalar product between spectra. Given the spectral invariance under unitary evolution (it can only permute the eigenvalues)~\cite{biswas2023}, Eq.~\eqref{Birkhoff1} implies
\begin{equation}
|\rho\rangle \rangle \succ  |\mathcal{D}(\rho^\prime)\rangle \rangle.
    \label{Birkhoff2}
\end{equation}
This majorization relation can be further interpreted via the Schur–Horn theorem~\cite{biswas2023,lobejko2024,biswas2024}, which ensures the existence of a bistochastic matrix $\Lambda$ such that
\begin{align}
  |\mathcal{D}(\rho^\prime)\rangle \rangle=\Lambda |\rho\rangle \rangle.
\end{align}
Therefore, it is straightforward to rewrite the expression of $Q_c$ in Eq.~\eqref{Qc_linear} as 
\begin{equation}
Q_c=\langle\langle H_c|\rho\rangle\rangle-\langle\langle H_c|\Lambda|\rho\rangle\rangle.
    \label{W_double}
\end{equation}
Bistochastic matrix will facilitates the transformation of $Q_c$ into terms involving permutation operations. From Birkhoff's theorem~\cite{sagawa2022,chen2019}, one can decompose any bistochastic matrix as a convex sum of permutation matrices $\Pi_m$ with coefficients $\alpha_m>0$ and $\sum_m\alpha_m=1$, i.e., $\Lambda = \sum_m \alpha_m \Pi_m$. Therefore, we have
\begin{align}
   Q_c&=\langle\langle H_c|\rho\rangle\rangle-\sum_m \alpha_m\langle\langle H_c|\Pi_m|\rho\rangle\rangle \nonumber \\
   &\le \langle\langle H_c|\rho\rangle\rangle-\underset{m}{\mathrm{min}}~  \langle\langle H_c|\Pi_m|\rho\rangle\rangle.  \label{extreme W}
\end{align}
It means that there exists a specific permutation that maximizes $Q_c$.

With the definition of $Q_c$, we can express the COP in Eq.~\eqref{COP} as follows:
\begin{align}
   \mathrm{COP}&=\frac{\sum _m \alpha_m Q^{\Pi_m}_c}{W} \nonumber \\
   &=\sum_m \frac{\alpha_m W^{\Pi_m}}{W} \frac{Q^{\Pi_m}_c}{W^{\Pi_m}}=\sum_m \frac{\alpha_m W^{\Pi_m}}{W} \mathrm{COP}^{\Pi_m}
       \label{COP_op}
\end{align}
where 
\begin{equation}
W^{\Pi_m}= \langle\langle (H_h+H_c)|\Pi_m|\rho\rangle\rangle-\langle\langle (H_h+H_c)|\rho\rangle\rangle,
    \label{W_pi}
\end{equation}

\begin{equation}
Q^{\Pi_m}_c=\langle\langle H_c|\rho\rangle\rangle- \langle\langle H_c|\Pi_m|\rho\rangle\rangle \nonumber,
\end{equation}
and 
\begin{equation}
\mathrm{COP}^{\Pi_m}=  \frac{Q^{\Pi_m}_c}{W^{\Pi_m}}.
\end{equation}
So far, we have reformulated the expression for the COP, which was previously represented merely as a formal symbol due to the broad scope of unitary operations, into a concrete summation of measurable physical quantities derived from a sequence of permutation operations. However, to obtain the bound for the COP, we must examine the sign of each term.

In fact, for the sake of argument all permutations involved in Eq.~\eqref{COP_op} must be the elements of one of the following sets~\cite{biswas2024}:
\begin{equation}
\mathcal{J}_R \equiv \{ m~\text{such that}~W^{\Pi_m}>0 ~\text{and} ~Q_c^{\Pi_m} >0\},
    \label{set_1}
\end{equation}
\begin{equation}
\mathcal{J}_E \equiv \{ m~\text{such that}~W^{\Pi_m}<0 ~\text{and}~ Q_c^{\Pi_m} <0\},
    \label{set_2}
\end{equation}
and
\begin{equation}
\mathcal{J}_A \equiv \{ m~\text{such that}~W^{\Pi_m}>0 ~\text{and} ~Q_c^{\Pi_m} \le 0\}.
    \label{set_3}
\end{equation}
For $m\in \mathcal{J}_R$, the permutation $\Pi_m$ corresponds to the model of a refrigerator. When  $m\in \mathcal{J}_E$ , the permutation $\Pi_m$ represents the mode of a heat engine. Similarly, $m\in \mathcal{J}_A$, the permutation $\Pi_m$ describes the mode of a heat accelerator. The condition $W^{\Pi_m}<0$ and ~$Q_c^{\Pi_m} \ge 0$ has been excluded from consideration as it violates the second law of thermodynamics. 

Thus, in order to determine the sign of each term, the COP given by Eq.~\eqref{COP_op} can be separated as 
\begin{align}
   \mathrm{COP}&=\sum_{m \in \mathcal{J}_R} \frac{\alpha_m W^{\Pi_m}}{W} \mathrm{COP}^{\Pi_m} \nonumber \\
   &+ \sum_{m \in \mathcal{J}_E} \frac{\alpha_m W^{\Pi_m}}{W} \mathrm{COP}^{\Pi_m}+\sum_{m \in \mathcal{J}_A} \frac{\alpha_m W^{\Pi_m}}{W} \mathrm{COP}^{\Pi_m}
   \label{COP_op_seprare}
\end{align}
Now, let us define $$\mathrm{COP}^{\mathrm{max/min}}_{\mathcal{J}_x}\equiv \underset{\Pi_m \in \mathcal{J}_x}{\mathrm{max/min}}~\mathrm{COP}^{\Pi_m},$$
 where $\mathrm{COP}_m^{\mathrm{max}}$ and $\mathrm{COP}_m^{\mathrm{min}}$  denotes the maximum and minimum $\mathrm{COP}$ when the two-stroke thermal machine operates in the mode $m$, respectively. Accordingly, we can express the following inequality: 
\begin{align}
   \mathrm{COP}&\le \sum_{m \in \mathcal{J}_R} \frac{\alpha_m W^{\Pi_m}}{W} \mathrm{COP}_{\mathcal{J}_R}^{\mathrm{max}} \nonumber \\
   &+ \sum_{m \in \mathcal{J}_E} \frac{\alpha_m W^{\Pi_m}}{W} \mathrm{COP}_{\mathcal{J}_E}^{\mathrm{min}}+\sum_{m \in \mathcal{J}_A} \frac{\alpha_m W^{\Pi_m}}{W}\mathrm{COP}_{\mathcal{J}_A}^{\mathrm{min}},
   \label{COP_op_inequality}
\end{align}
where we use the fact $\frac{\alpha_m W^{\Pi_m}}{W} \mathrm{COP}^{\Pi_m}$>0 for all $m\in\mathcal{J}_R$, whereas $\frac{\alpha_m W^{\Pi_m}}{W} \mathrm{COP}^{\Pi_m} < 0$ for all $m\in\mathcal{J}_E$ and  $m\in\mathcal{J}_A$, as defined in Eqs.~\eqref{set_1}-\eqref{set_3}. Note that in a heat accelerator, the energy from external work is completely converted into heat without yielding any useful energy output, which represents a meaningless or inefficient mode of operation. 

In order to further get the bound of COP, we need to establish the order of COP among various operating modes within the allowable range of the laws of thermodynamics. From Eq.~\eqref{2nd_prove_2}, we can easily get the bound for any refrigerator as $0< \mathrm{COP}_{\mathrm{refrigerator}} \le \beta_h/(\beta_c-\beta_h) \equiv \mathrm{COP}_{\mathrm{Carnot}}$. Eq.~\eqref{2nd_prove_2}  also indicates that any two-stroke thermal machine in the engine mode always exhibits a negative $Q_c$ and $Q_h$, as
\begin{equation}
0>W=Q_h-Q_c=Q_c(\frac{Q_h}{Q_c}-1)\geq Q_c(\frac{\beta_c}{\beta_h}-1),
    \label{JN}
\end{equation}
where $\beta_c/\beta_h > 1$. This also clarifies the rationale behind dividing the data into four sets as engine, refrigerator and accelerator. Furthermore, by employing Eqs.~\eqref{2nd_prove_2} and \eqref{JN}, we can establish the ordering of the COP among the different operational modes of the two-stroke thermal machines. 

For a two-stroke thermal machine operating in the engine mode, we have $Q_c < 0$. Therefore,  the second law inequality given in Eq.~\eqref{2nd_prove_2} can be reduced to $Q_h/Q_c \le \frac{\beta_c}{\beta_h}$, which implies $\mathrm{COP}_{\mathrm{engine}}=(Q_h/Q_c-1)^{-1} \ge (\beta_c/\beta_h-1)^{-1}$ and $\mathrm{COP}^{\mathrm{min}}_{\mathrm{engine}}=\mathrm{min}[(Q_h/Q_c-1)^{-1}] \ge (\beta_c/\beta_h-1)^{-1} $

Then, for a two-stroke thermal machine operating in the refrigerator mode, we have $\mathrm{COP}^{\mathrm{min}}_{\mathrm{refrigerator}}=\mathrm{min}[(Q_h/Q_c-1)^{-1}]$ and $\mathrm{COP}^{\mathrm{max}}_{\mathrm{refrigerator}}=\mathrm{max}[(Q_h/Q_c-1)^{-1}]$. Both of them are in the range $[0,(\beta_c/\beta_h-1)^{-1}]$.

Finally, for a two-stroke thermal machine operating as a heat accelerator, we have  $W >0$ and $Q_c \le 0$, which makes its $\mathrm{COP}_{\mathrm{heat~accelerators}}$ non-positive, i.e., $\mathrm{COP}_{\mathrm{heat~accelerator}}=(Q_h/Q_c-1)^{-1} \le 0$ thus $\mathrm{COP}^{\mathrm{max}}_{\mathrm{heat~accelerator}}=\mathrm{max}[(Q_h/Q_c-1)^{-1}] \le 0$. The second-law inequality in Eq.~\eqref{2nd_prove_2} establishes the COP ordering for the thermal machine’s distinct operational modes, which is demonstrated in Fig.~\ref{fig:COP_order} and mathematically expressed as
\begin{equation}
\mathrm{COP}^{\mathrm{max}}_{\mathrm{heat~accelerators}}\le \mathrm{COP}^{\mathrm{min}}_{\mathrm{refrigerator}} \le \mathrm{COP}^{\mathrm{max}}_{\mathrm{refrigerator}} \le \mathrm{COP}^{\mathrm{min}}_{\mathrm{engine}},
    \label{order}
\end{equation}
Then, with the above inequality, which is true for any unitaries that transforms the state of the thermal machine during work stroke, we can write
\begin{align}
   \mathrm{COP}^{\mathrm{max}}_{\mathcal{J}_A}&\le \mathrm{COP}^{\mathrm{max}}_{\mathrm{heat~accelerator}}\le \mathrm{COP}^{\mathrm{min}}_{\mathrm{refrigerator}} \nonumber \\
   &\le \mathrm{COP}^{\mathrm{max}}_{\mathcal{J}_R} \le \mathrm{COP}^{\mathrm{min}}_{\mathrm{engine}} \le \mathrm{COP}^{\mathrm{min}}_{\mathcal{J}_E}.
   \label{order_1}
\end{align}
Here, $\mathrm{COP}^{\mathrm{max}}_{\mathrm{heat~accelerator}}$,  $\mathrm{COP}^{\mathrm{min}}_{\mathrm{refrigerator}}$, and $\mathrm{COP}^{\mathrm{min}}_{\mathrm{engine}}$ correspond to the the maximum or minimum $\mathrm{COP}$ of global transformation of the mode of the heat accelerator, refrigerator, engine, respectively. Then, with Eq.~\eqref{order_1} and $W^{\Pi_m}<0$, we have
\begin{equation}
\frac{\alpha_m W^{\Pi_m}}{W}\mathrm{COP}^{\mathrm{min}}_{\mathcal{J}_E} \le \frac{\alpha_m W^{\Pi_m}}{W}\mathrm{COP}^{\mathrm{max}}_{\mathcal{J}_R}
    \label{E_inequality}
\end{equation}
for $m \in \mathcal{J}_E$. Combing Eq.~\eqref{order_1} with $W^{\Pi_m}>0$, we have
\begin{equation}
\frac{\alpha_m W^{\Pi_m}}{W}\mathrm{COP}^{\mathrm{max}}_{\mathcal{J}_A} \le \frac{\alpha_m W^{\Pi_m}}{W}\mathrm{COP}^{\mathrm{max}}_{\mathcal{J}_R},
    \label{A_inequality}
\end{equation}
for $m \in \mathcal{J}_A$. Substituting the above two inequalities into Eq.~\eqref{COP_op_inequality} yields
\begin{align}
   \mathrm{COP}&\le \sum_{m \in \mathcal{J}_R} \frac{\alpha_m W^{\Pi_m}}{W} \mathrm{COP}_{\mathcal{J}_R}^{\mathrm{max}} \nonumber \\
   &+ \sum_{m \in \mathcal{J}_E} \frac{\alpha_m W^{\Pi_m}}{W} \mathrm{COP}_{\mathcal{J}_R}^{\mathrm{max}}+\sum_{m \in \mathcal{J}_A} \frac{\alpha_m W^{\Pi_m}}{W}\mathrm{COP}_{\mathcal{J}_R}^{\mathrm{max}}\nonumber \\
   &\le \mathrm{COP}_{\mathcal{J}_R}^{\mathrm{max}}(\sum_m \frac{\alpha_m W^{\Pi_m}}{W})=\mathrm{COP}_{\mathcal{J}_R}^{\mathrm{max}},
   \label{COP_op_inequality_final}
\end{align}
which completes the proof.

\section{The calculation for 24 possible permutations of \textbf{$Q_c$} and $\mathrm{\textbf{COP}}$}
\label{appendix B}
The values of $Q_c$ and $\mathrm{COP}$ for 24 possible permutations of a noncatalytic two-stroke refrigerator are listed in Table~\ref{tab:table1}.
\begin{table*}
\caption{\label{tab:table1}
 The heat absorbed from the cold reservoir and the COP of the refrigerator when the working body composed of two TLSs that have been thermalized at different temperatures and subsequently transformed by all possible permutations. From the table, we observe that out of the 24 permutations, only four yield a positive value of  $Q_c$. Here, $a_{h/c} \equiv e^{-\beta_{h/c} \omega_{h/c}}$,  and a necessary condition for  $Q_c>0$ is  that  $a_c>a_h$.}
\begin{ruledtabular}
\begin{tabular}{cccc}
&Permutation matrices&
$Q_c$&
COP\\
\colrule
1 & $\mathbb{I}= |00\rangle\langle 00|+|01\rangle\langle 01|+|10\rangle\langle 10|+|11\rangle\langle 11|$ & $0$ & $0$\\
2 & $\Pi= |00\rangle\langle 00|+|01\rangle\langle 01|+|10\rangle\langle 11|+|11\rangle\langle 10|$ & $-\frac{a_h(1-a_c)\omega_c}{(1+a_h)(1+a_c)}$ & $-1$\\
3 & $|00\rangle\langle 00|+|01\rangle\langle 10|+|10\rangle\langle 01|+|11\rangle\langle 11|$ & $\frac{(a_c-a_h)\omega_c}{(1+a_h)(1+a_c)}$ & $\frac{\omega_c}{\omega_h-\omega_c}$\\
4 & $|00\rangle\langle 00|+|01\rangle\langle 10|+|10\rangle\langle 11|+|11\rangle\langle 01|$ & $-\frac{a_h(1-a_c)\omega_c}{(1+a_h)(1+a_c)}$ & $-\frac{\omega_c}{\frac{a_c-a_h}{a_h(1-a_c)}\omega_h +\omega_c}$\\
5 & $|00\rangle\langle 00|+|01\rangle\langle 11|+|10\rangle\langle 01|+|11\rangle\langle 10|$ & $\frac{(a_c-a_h)\omega_c}{(1+a_h)(1+a_c)}$ & $\frac{\omega_c}{\frac{1-a_h}{1-a_h/a_c}\omega_h-\omega_c}$\\
6 & $|00\rangle\langle 00|+|01\rangle\langle 11|+|10\rangle\langle 10|+|11\rangle\langle 01|$ & $0$ & $0$\\
7 & $|00\rangle\langle 10|+|01\rangle\langle 00|+|10\rangle\langle 10|+|11\rangle\langle 11|$ & $-\frac{(1-a_c)\omega_c}{(1+a_h)(1+a_c)}$ & $-1$\\
8 & $|00\rangle\langle 01|+|01\rangle\langle 00|+|10\rangle\langle 11|+|11\rangle\langle 10|$ & $-\frac{(1-a_c)\omega_c}{1+a_c}$ & $-1$\\
9 & $|00\rangle\langle 01|+|01\rangle\langle 10|+|10\rangle\langle 00|+|11\rangle\langle 11|$ & $\frac{(a_c-a_h)\omega_c}{(1+a_h)(1+a_c)}$ & $\frac{\omega_c}{\frac{1-a_h}{a_c-a_h}\omega_h-\omega_c}$\\
10 & $|00\rangle\langle 01|+|01\rangle\langle 10|+|10\rangle\langle 11|+|11\rangle\langle 00|$ & $-\frac{(1-a_c)\omega_c}{1+a_c}$ & $-\frac{\omega_c}{\frac{1-a_h}{(1-a_c)(1+a_h)}\omega_h+\omega_c}$\\
11 & $|00\rangle\langle 01|+|01\rangle\langle 11|+|10\rangle\langle 00|+|11\rangle\langle 10|$ & $\frac{(a_c-a_h)\omega_c}{(1+a_h)(1+a_c)}$ & $\frac{\omega_c}{\frac{1-a_ca_h}{a_c-a_h}\omega_h-\omega_c}$\\
12 & $|00\rangle\langle 01|+|01\rangle\langle 11|+|10\rangle\langle 10|+|11\rangle\langle 00|$ & $-\frac{(1-a_c)\omega_c}{(1+a_h)(1+a_c)}$ & $-\frac{\omega_c}{\frac{1-a_ca_h}{1-a_c}\omega_h+\omega_c}$\\
13 & $|00\rangle\langle 10|+|01\rangle\langle 00|+|10\rangle\langle 01|+|11\rangle\langle 11|$ & $-\frac{(1-a_c)\omega_c}{(1+a_h)(1+a_c)}$ & $-\frac{\omega_c}{\frac{a_c-a_h}{1-a_c}\omega_h+\omega_c}$\\
14 & $|00\rangle\langle 10|+|01\rangle\langle 00|+|10\rangle\langle 11|+|11\rangle\langle 01|$ & $-\frac{(1-a_ca_h)\omega_c}{(1+a_h)(1+a_c)}$ & $-\frac{\omega_c}{\frac{a_c-a_h}{1-a_c a_h}\omega_h+\omega_c}$\\
15 & $|00\rangle\langle 10|+|01\rangle\langle 01|+|10\rangle\langle 00|+|11\rangle\langle 11|$ & $0$ & $0$\\
16 & $|00\rangle\langle 10|+|01\rangle\langle 01|+|10\rangle\langle 11|+|11\rangle\langle 00|$ & $-\frac{(1-a_ca_h)\omega_c}{(1+a_h)(1+a_c)}$ & $-\frac{\omega_c}{\frac{1-a_h}{1-a_ca_h}\omega_h+\omega_c}$\\
17 & $|00\rangle\langle 10|+|01\rangle\langle 11|+|10\rangle\langle 00|+|11\rangle\langle 01|$ & $0$ & $0$\\
18 & $|00\rangle\langle 10|+|01\rangle\langle 11|+|10\rangle\langle 01|+|11\rangle\langle 00|$ & $-\frac{(1-a_c)\omega_c}{(1+a_h)(1+a_c)}$ & $-\frac{\omega_c}{\frac{(1-a_h)(1+a_c)}{1-a_c}\omega_h+\omega_c}$\\
19 & $|00\rangle\langle 11|+|01\rangle\langle 00|+|10\rangle\langle 01|+|11\rangle\langle 10|$ & $-\frac{(1-a_c)\omega_c}{1+a_c}$ & $-\frac{\omega_c}{\frac{a_c(1-a_h)}{(1-a_c)(1+a_h)}\omega_h+\omega_c}$\\
20 & $|00\rangle\langle 11|+|01\rangle\langle 00|+|10\rangle\langle 10|+|11\rangle\langle 01|$ & $-\frac{(1-a_ca_h)\omega_c}{(1+a_h)(1+a_c)}$ & $-\frac{\omega_c}{\frac{a_c(1-a_h)}{1-a_ca_h}\omega_h+\omega_c}$\\
21 & $|00\rangle\langle 11|+|01\rangle\langle 01|+|10\rangle\langle 00|+|11\rangle\langle 10|$ & $-\frac{a_h(1-a_c)\omega_c}{(1+a_h)(1+a_c)}$ & $-\frac{\omega_c}{\frac{1-a_ca_h}{a_h(1-a_c)}\omega_h+\omega_c}$\\
22 & $|00\rangle\langle 11|+|01\rangle\langle 01|+|10\rangle\langle 10|+|11\rangle\langle 00|$ & $-\frac{(1-a_ca_h)\omega_c}{(1+a_h)(1+a_c)}$ & $-\frac{\omega_c}{\omega_h+\omega_c}$\\
23 & $|00\rangle\langle 11|+|01\rangle\langle 10|+|10\rangle\langle 00|+|11\rangle\langle 01|$ & $-\frac{a_h(1-a_c)\omega_c}{(1+a_h)(1+a_c)}$ & $-\frac{\omega_c}{\frac{(1+a_c)(1-a_h)}{a_h(1-a_c)}\omega_h+\omega_c}$\\
24 & $|00\rangle\langle 11|+|01\rangle\langle 10|+|10\rangle\langle 01|+|11\rangle\langle 00|$ & $-\frac{(1-a_c)\omega_c}{1+a_c}$ & $-\frac{\omega_c}{\frac{(1+a_c)(1-a_h)}{(1-a_c)(1+a_h)}\omega_h+\omega_c}$\\
\end{tabular}
\end{ruledtabular}
\end{table*}

\section{Detailed derivations of $\delta P$}
\label{appendix C}
The derivation proceeds by first establishing the constraints for catalyst preservation
\begin{align}
   \delta P&=\mathcal{N}(p_{m+1}a_c-p_m )\quad \mathrm{for}~m\in \{ 1,2,\cdots,n \}, \nonumber\\
   \delta P&=\mathcal{N}(p_{n+t+1}a_c-p_{n+t} a_h )\quad \mathrm{for}~t\in \{ 1,2,\cdots,n^{\prime}-1 \},\nonumber
\end{align}
and
\begin{equation}
\delta P=\mathcal{N}(p_{1}a_c-p_{n+n^{\prime}} a_h ),
\label{conditions for preserving the catalyst}
\end{equation}
where $$\mathcal{N}=\frac{1}{(1+a_h)(1+a_c)}.$$ With Eq.~\eqref{conditions for preserving the catalyst}, we have
\begin{equation}
p_{m+1}=\frac{p_m}{a_c}+\frac{\delta P}{\mathcal{N} a_c}\quad \mathrm{for}~m\in \{ 1,2,\cdots,n \},
\label{Eq39_1}
\end{equation}
\begin{equation}
p_{n+m^\prime+1}=p_{n+m^\prime}\frac{a_h}{a_c}+\frac{\delta P}{\mathcal{N}a_c}\quad \mathrm{for}~m^\prime\in \{ 1,2,\cdots,n^{\prime}-1 \},
\label{Eq39_2}
\end{equation}
and
\begin{equation}
p_{1}=p_{n+n^{\prime}}\frac{a_h}{a_c}+\frac{\delta P}{\mathcal{N}a_c},
\label{Eq39_3}
\end{equation}
which can be further simplified as
\begin{align}
    \nonumber\\
   p_{m+1}&=\frac{p_m}{a_c}+\frac{\delta P}{\mathcal{N} a_c}\nonumber\\
   &=\frac{p_1}{a_c^m}+\frac{\delta P}{\mathcal{N} a_c^m}+\frac{\delta P}{\mathcal{N} a_c^{m-1}}+\cdots+\frac{\delta P}{\mathcal{N} a_c}\nonumber\\
   &=\frac{p_1}{a_c^m}+\frac{\delta P}{\mathcal{N}}\frac{a_c^m-1}{a_c^{m+1}-a_c^m} ~\mathrm{for}~m\in \{ 1,2,\cdots,n \},
   \label{Eq39_1_sim}
\end{align}
\begin{align}
    \nonumber\\
   p_{n+m^\prime+1}&=p_{n+m^\prime}\frac{a_h}{a_c}+\frac{\delta P}{\mathcal{N}a_c}\nonumber\\
   &=(\frac{p_1}{a_c^n}+\frac{\delta P}{\mathcal{N}}\frac{a_c^n-1}{a_c^{n+1}-a_c^n}) (\frac{a_h}{a_c})^{m^\prime }\nonumber\\
   &+\frac{\delta P}{\mathcal{N}a_c} \left[\frac{1-(\frac{a_h}{a_c})^{m^\prime}}{1-\frac{a_h}{a_c}}\right] ~\mathrm{for}~m^\prime\in \{ 1,2,\cdots,n^{\prime}-1 \},
   \label{Eq39_2_sim}
\end{align}
and
\begin{equation}
p_{1}=p_{n+n^{\prime}}\frac{a_h}{a_c}+\frac{\delta P}{\mathcal{N}a_c}.
\label{Eq39_3_sim}
\end{equation}
From the normalization of the probability, we can write
\begin{align}
\sum_{x=2}^{n+n^{\prime}}p_x&=\sum_{x=2}^{n+1}p_{x}+\sum_{x=n+2}^{n+n^{\prime}}p_{x}=\sum_{m=1}^{n}p_{m+1}+\sum_{m^\prime=1}^{n^{\prime}-1}p_{n+m^\prime+1}\nonumber\\
&=1-p_1.
\label{normalized probability}
\end{align}
Substituting the values of $p_{m+1}$ and $p_{n+m^\prime+1}$ from Eq.~\eqref{Eq39_1_sim} and Eq.~\eqref{Eq39_2_sim} in Eq.~\eqref{normalized probability}, we obtain $p_1$ in terms of $\delta P$. Then, with another relation between $p_1$ and $\delta P$ in Eq.~\eqref{Eq39_3_sim}, we have Eq.~\eqref{delta_P} with
\begin{widetext}
\begin{equation}
f(a_c,a_h,n,n^{\prime}) \equiv \frac{a_{c}\left(1-a_{h}\right)^{2}\left\{\left(1-a_{c}^{n}\right)\left(a_{c}^{n^{\prime}}-a_{h}^{n^{\prime}}\right)\right\}+\left\{\left(a_{c}^{(n+n^{\prime})}-a_{h}^{n^{\prime}}\right)\left(a_{c}-a_{h}\right)\left(1-a_{c}\right)\right\}\left\{n^{\prime}\left(1-a_{c}\right)-n\left(a_{c}-a_{h}\right)\right\}}{(1-a_c)^2(a_c-a_h)^2}. 
\label{delta_P_f}
\end{equation}
\end{widetext}

\bibliography{main}

@article{Von2019,
  title = {Von {{Neumann Entropy}} from {{Unitarity}}},
  author = {Boes, Paul and Eisert, Jens and Gallego, Rodrigo and M{\"u}ller, Markus P. and Wilming, Henrik},
  year = {2019},
  month = may,
  journal = {Phys. Rev. Lett.},
  volume = {122},
  number = {21},
  pages = {210402},
  issn = {0031-9007, 1079-7114},
  doi = {10.1103/PhysRevLett.122.210402},
  urldate = {2025-04-02}
}

@article{brand2015,
  title = {The Second Laws of Quantum Thermodynamics},
  author = {Brand{\~a}o, Fernando and Horodecki, Micha{\l} and Ng, Nelly and Oppenheim, Jonathan and Wehner, Stephanie},
  year = {2015},
  month = mar,
  journal = {Proc. Natl. Acad. Sci. U. S. A.},
  volume = {112},
  number = {11},
  pages = {3275--3279},
  issn = {0027-8424, 1091-6490},
  doi = {10.1073/pnas.1411728112},
  urldate = {2025-05-16},
  langid = {english}
}

@article{aberg2014,
  title = {Catalytic {{Coherence}}},
  author = {{\AA}berg, Johan},
  year = {2014},
  month = oct,
  journal = {Phys. Rev. Lett.},
  volume = {113},
  number = {15},
  pages = {150402},
  issn = {0031-9007, 1079-7114},
  doi = {10.1103/PhysRevLett.113.150402},
  urldate = {2025-05-18},
  copyright = {http://link.aps.org/licenses/aps-default-license},
  langid = {english}
}

@article{ng2015,
  title = {Limits to Catalysis in Quantum Thermodynamics},
  author = {Ng, N H Y and Man{\v c}inska, L and Cirstoiu, C and Eisert, J and Wehner, S},
  year = {2015},
  month = aug,
  journal = {New J. Phys.},
  volume = {17},
  number = {8},
  pages = {085004},
  issn = {1367-2630},
  doi = {10.1088/1367-2630/17/8/085004},
  urldate = {2025-05-15},
  langid = {english}
}

@article{lostaglio2015,
  title = {Stochastic {{Independence}} as a {{Resource}} in {{Small-Scale Thermodynamics}}},
  author = {Lostaglio, Matteo and M{\"u}ller, Markus P. and Pastena, Michele},
  year = {2015},
  month = oct,
  journal = {Phys. Rev. Lett.},
  volume = {115},
  number = {15},
  pages = {150402},
  issn = {0031-9007, 1079-7114},
  doi = {10.1103/PhysRevLett.115.150402},
  urldate = {2025-05-18},
  copyright = {http://link.aps.org/licenses/aps-default-license},
  langid = {english}
}

@article{muller2018,
  title = {Correlating {{Thermal Machines}} and the {{Second Law}} at the {{Nanoscale}}},
  author = {M{\"u}ller, Markus P.},
  year = {2018},
  month = dec,
  journal = {Phys. Rev. X},
  volume = {8},
  number = {4},
  pages = {041051},
  issn = {2160-3308},
  doi = {10.1103/PhysRevX.8.041051},
  urldate = {2025-05-18},
  langid = {english}
}

@article{shiraishi2021,
  title = {Quantum {{Thermodynamics}} of {{Correlated-Catalytic State Conversion}} at {{Small Scale}}},
  author = {Shiraishi, Naoto and Sagawa, Takahiro},
  year = {2021},
  month = apr,
  journal = {Phys. Rev. Lett.},
  volume = {126},
  number = {15},
  pages = {150502},
  issn = {0031-9007, 1079-7114},
  doi = {10.1103/PhysRevLett.126.150502},
  urldate = {2025-05-18},
  langid = {english}
}

@article{biswas2024,
  title = {Catalytic Enhancement in the Performance of the Microscopic Two-Stroke Heat Engine},
  author = {Biswas, Tanmoy and {\L}obejko, Marcin and Mazurek, Pawe{\l} and Horodecki, Micha{\l}},
  year = {2024},
  month = oct,
  journal = {Phys. Rev. E},
  volume = {110},
  number = {4},
  pages = {044120},
  issn = {2470-0045, 2470-0053},
  doi = {10.1103/PhysRevE.110.044120},
  urldate = {2025-01-13},
  langid = {english}
}

@article{henao2023,
  title = {Catalytic {{Leverage}} of {{Correlations}} and {{Mitigation}} of {{Dissipation}} in {{Information Erasure}}},
  author = {Henao, I. and Uzdin, R.},
  year = {2023},
  month = jan,
  journal = {Phys. Rev. Lett.},
  volume = {130},
  number = {2},
  pages = {020403},
  issn = {0031-9007, 1079-7114},
  doi = {10.1103/PhysRevLett.130.020403},
  urldate = {2025-03-06},
  langid = {english}
}

@article{henao2021,
  title = {Catalytic Transformations with Finite-Size Environments: Applications to Cooling and Thermometry},
  shorttitle = {Catalytic Transformations with Finite-Size Environments},
  author = {Henao, Ivan and Uzdin, Raam},
  year = {2021},
  month = sep,
  journal = {Quantum},
  volume = {5},
  pages = {547},
  doi = {10.22331/q-2021-09-21-547},
  urldate = {2025-05-09},
  langid = {english}
}

@article{lobejko2024,
  title = {Catalytic {{Advantage}} in {{Otto-like Two-Stroke Quantum Engines}}},
  author = {{\L}obejko, Marcin and Biswas, Tanmoy and Mazurek, Pawe{\l} and Horodecki, Micha{\l}},
  year = {2024},
  month = jun,
  journal = {Phys. Rev. Lett.},
  volume = {132},
  number = {26},
  pages = {260403},
  issn = {0031-9007, 1079-7114},
  doi = {10.1103/PhysRevLett.132.260403},
  urldate = {2024-07-24},
  langid = {english}
}

@article{sparaciari2017,
  title = {Energetic Instability of Passive States in Thermodynamics},
  author = {Sparaciari, Carlo and Jennings, David and Oppenheim, Jonathan},
  year = {2017},
  month = dec,
  journal = {Nat. Commun.},
  volume = {8},
  number = {1},
  pages = {1895},
  issn = {2041-1723},
  doi = {10.1038/s41467-017-01505-4},
  urldate = {2025-05-18},
  langid = {english}
}

@article{wilming2021,
  title = {Entropy and {{Reversible Catalysis}}},
  author = {Wilming, H.},
  year = {2021},
  month = dec,
  journal = {Phys. Rev. Lett.},
  volume = {127},
  number = {26},
  pages = {260402},
  issn = {0031-9007, 1079-7114},
  doi = {10.1103/PhysRevLett.127.260402},
  urldate = {2025-05-18},
  langid = {english}
}

@article{jonathan1999,
  title = {Entanglement-{{Assisted Local Manipulation}} of {{Pure Quantum States}}},
  author = {Jonathan, Daniel and Plenio, Martin B.},
  year = {1999},
  month = oct,
  journal = {Phys. Rev. Lett.},
  volume = {83},
  number = {17},
  pages = {3566--3569},
  issn = {0031-9007, 1079-7114},
  doi = {10.1103/PhysRevLett.83.3566},
  urldate = {2025-05-18},
  copyright = {http://link.aps.org/licenses/aps-default-license},
  langid = {english}
}

@article{anshu2018,
  title = {Quantifying {{Resources}} in {{General Resource Theory}} with {{Catalysts}}},
  author = {Anshu, Anurag and Hsieh, Min-Hsiu and Jain, Rahul},
  year = {2018},
  month = nov,
  journal = {Phys. Rev. Lett.},
  volume = {121},
  number = {19},
  pages = {190504},
  issn = {0031-9007, 1079-7114},
  doi = {10.1103/PhysRevLett.121.190504},
  urldate = {2025-05-30},
  langid = {english}
}

@article{boes2018,
  title = {Catalytic {{Quantum Randomness}}},
  author = {Boes, P. and Wilming, H. and Gallego, R. and Eisert, J.},
  year = {2018},
  month = oct,
  journal = {Phys. Rev. X},
  volume = {8},
  number = {4},
  pages = {041016},
  issn = {2160-3308},
  doi = {10.1103/PhysRevX.8.041016},
  urldate = {2025-05-30},
  langid = {english}
}

@article{daftuar2001,
  title = {Mathematical Structure of Entanglement Catalysis},
  author = {Daftuar, Sumit and Klimesh, Matthew},
  year = {2001},
  month = sep,
  journal = {Phys. Rev. A},
  volume = {64},
  number = {4},
  pages = {042314},
  issn = {1050-2947, 1094-1622},
  doi = {10.1103/PhysRevA.64.042314},
  urldate = {2025-05-30},
  copyright = {http://link.aps.org/licenses/aps-default-license},
  langid = {english}
}

@article{ding2021,
  title = {Amplifying Asymmetry with Correlating Catalysts},
  author = {Ding, Feng and Hu, Xueyuan and Fan, Heng},
  year = {2021},
  month = feb,
  journal = {Phys. Rev. A},
  volume = {103},
  number = {2},
  pages = {022403},
  issn = {2469-9926, 2469-9934},
  doi = {10.1103/PhysRevA.103.022403},
  urldate = {2025-05-30},
  langid = {english}
}

@article{majenz2017,
  title = {Catalytic {{Decoupling}} of {{Quantum Information}}},
  author = {Majenz, Christian and Berta, Mario and Dupuis, Fr{\'e}d{\'e}ric and Renner, Renato and Christandl, Matthias},
  year = {2017},
  month = feb,
  journal = {Phys. Rev. Lett.},
  volume = {118},
  number = {8},
  pages = {080503},
  issn = {0031-9007, 1079-7114},
  doi = {10.1103/PhysRevLett.118.080503},
  urldate = {2025-05-30},
  copyright = {http://link.aps.org/licenses/aps-default-license},
  langid = {english}
}

@article{rethinasamy2020,
  title = {Relative Entropy and Catalytic Relative Majorization},
  author = {Rethinasamy, Soorya and Wilde, Mark M.},
  year = {2020},
  month = sep,
  journal = {Phys. Rev. Res.},
  volume = {2},
  number = {3},
  pages = {033455},
  issn = {2643-1564},
  doi = {10.1103/PhysRevResearch.2.033455},
  urldate = {2025-05-30},
  langid = {english}
}

@article{sandersNecessary2009,
  title = {Necessary Conditions for Entanglement Catalysts},
  author = {Sanders, Yuval Rishu and Gour, Gilad},
  year = {2009},
  month = may,
  journal = {Phys. Rev. A},
  volume = {79},
  number = {5},
  pages = {054302},
  issn = {1050-2947, 1094-1622},
  doi = {10.1103/PhysRevA.79.054302},
  urldate = {2025-05-30},
  copyright = {http://link.aps.org/licenses/aps-default-license}
}

@article{turgut2007,
  title = {Catalytic Transformations for Bipartite Pure States},
  author = {Turgut, S},
  year = {2007},
  month = oct,
  journal = {J. Phys. A-Math. Theor.},
  volume = {40},
  number = {40},
  pages = {12185--12212},
  issn = {1751-8113, 1751-8121},
  doi = {10.1088/1751-8113/40/40/012},
  langid = {english}
}

@article{allahverdyan2011,
  title = {Work Extraction from Microcanonical Bath},
  author = {Allahverdyan, A. E. and Hovhannisyan, K. V.},
  year = {2011},
  month = sep,
  journal = {Europhys. Lett.},
  volume = {95},
  number = {6},
  pages = {60004},
  issn = {0295-5075, 1286-4854},
  doi = {10.1209/0295-5075/95/60004},
  urldate = {2025-05-30},
  langid = {english}
}

@article{boes2020,
  title = {By-Passing Fluctuation Theorems},
  author = {Boes, Paul and Gallego, Rodrigo and Ng, Nelly H. Y. and Eisert, Jens and Wilming, Henrik},
  year = {2020},
  month = feb,
  journal = {Quantum},
  volume = {4},
  eprint = {1904.01314},
  pages = {231},
  doi = {10.22331/q-2020-02-20-231},
  urldate = {2025-05-30},
  langid = {english}
}

@article{lipka-bartosik2021a,
  title = {All {{States}} Are {{Universal Catalysts}} in {{Quantum Thermodynamics}}},
  author = {{Lipka-Bartosik}, Patryk and Skrzypczyk, Paul},
  year = {2021},
  month = mar,
  journal = {Phys. Rev. X},
  volume = {11},
  number = {1},
  pages = {011061},
  issn = {2160-3308},
  doi = {10.1103/PhysRevX.11.011061},
  urldate = {2025-05-30},
  langid = {english}
}

@article{wilming2017,
  title = {Third {{Law}} of {{Thermodynamics}} as a {{Single Inequality}}},
  author = {Wilming, Henrik and Gallego, Rodrigo},
  year = {2017},
  month = nov,
  journal = {Phys. Rev. X},
  volume = {7},
  number = {4},
  pages = {041033},
  issn = {2160-3308},
  doi = {10.1103/PhysRevX.7.041033},
  urldate = {2025-05-30},
  copyright = {https://creativecommons.org/licenses/by/4.0/},
  langid = {english}
}

@book{bhatia1997,
  title = {Matrix {{Analysis}}},
  author = {Bhatia, Rajendra},
  year = {1997},
  series = {Graduate {{Texts}} in {{Mathematics}}},
  volume = {169},
  publisher = {Springer New York},
  address = {New York, NY},
  doi = {10.1007/978-1-4612-0653-8},
  urldate = {2025-05-30},
  copyright = {http://www.springer.com/tdm},
  isbn = {978-1-4612-6857-4 978-1-4612-0653-8},
  langid = {english}
}

@article{D.Blackwell1953,
  title = {Equivalent Comparisons of Experiments},
  author = {David Blackwell},
  year = {1953},
  journal = {Ann. Math. Stat.},
  volume = {24},
  pages = {265},
  doi = {10.1214/aoms/1177729032}
}

@article{clivaz2019,
  title = {Unifying {{Paradigms}} of {{Quantum Refrigeration}}: {{A Universal}} and {{Attainable Bound}} on {{Cooling}}},
  author = {Clivaz, Fabien and Silva, Ralph and Haack, G{\'e}raldine and Brask, Jonatan Bohr and Brunner, Nicolas and Huber, Marcus},
  year = {2019},
  month = oct,
  journal = {Phys. Rev. Lett.},
  volume = {123},
  number = {17},
  pages = {170605},
  issn = {0031-9007, 1079-7114},
  doi = {10.1103/PhysRevLett.123.170605},
  urldate = {2025-03-19}
}

@article{Brunner2014,
  title = {Entanglement enhances cooling in microscopic quantum refrigerators},
  author = {Brunner, Nicolas and Huber, Marcus and Linden, Noah and Popescu, Sandu and Silva, Ralph and Skrzypczyk, Paul},
  journal = {Phys. Rev. E},
  volume = {89},
  issue = {3},
  pages = {032115},
  numpages = {6},
  year = {2014},
  month = {Mar},
  publisher = {American Physical Society},
  doi = {10.1103/PhysRevE.89.032115},
  url = {https://link.aps.org/doi/10.1103/PhysRevE.89.032115}
}

@article{Correa2014,
  title = {Optimal performance of endoreversible quantum refrigerators},
  author = {Correa, Luis A. and Palao, Jos\'e P. and Adesso, Gerardo and Alonso, Daniel},
  journal = {Phys. Rev. E},
  volume = {90},
  issue = {6},
  pages = {062124},
  numpages = {7},
  year = {2014},
  month = {Dec},
  publisher = {American Physical Society},
  doi = {10.1103/PhysRevE.90.062124},
  url = {https://link.aps.org/doi/10.1103/PhysRevE.90.062124}
}

@article{kolar2012,
  title = {Quantum {{Bath Refrigeration}} towards {{Absolute Zero}}: {{Challenging}} the {{Unattainability Principle}}},
  shorttitle = {Quantum {{Bath Refrigeration}} towards {{Absolute Zero}}},
  author = {Kol{\'a}{\v r}, M. and {Gelbwaser-Klimovsky}, D. and Alicki, R. and Kurizki, G.},
  year = {2012},
  month = aug,
  journal = {Phys. Rev. Lett.},
  volume = {109},
  number = {9},
  pages = {090601},
  issn = {0031-9007, 1079-7114},
  doi = {10.1103/PhysRevLett.109.090601},
  urldate = {2025-05-30},
  copyright = {http://link.aps.org/licenses/aps-default-license},
  langid = {english}
}

@article{levy2012a,
  title = {Quantum {{Absorption Refrigerator}}},
  author = {Levy, Amikam and Kosloff, Ronnie},
  year = {2012},
  month = feb,
  journal = {Phys. Rev. Lett.},
  volume = {108},
  number = {7},
  pages = {070604},
  issn = {0031-9007, 1079-7114},
  doi = {10.1103/PhysRevLett.108.070604},
  urldate = {2025-05-30},
  copyright = {http://link.aps.org/licenses/aps-default-license},
  langid = {english}
}

@article{soldati2022,
  title = {Thermodynamics of a {{Minimal Algorithmic Cooling Refrigerator}}},
  author = {Soldati, Rodolfo R. and Dasari, Durga B. R. and Wrachtrup, J{\"o}rg and Lutz, Eric},
  year = {2022},
  month = jul,
  journal = {Phys. Rev. Lett.},
  volume = {129},
  number = {3},
  pages = {030601},
  issn = {0031-9007, 1079-7114},
  doi = {10.1103/PhysRevLett.129.030601},
  urldate = {2025-05-30},
  langid = {english}
}

@article{venturelli2013,
  title = {Minimal {{Self-Contained Quantum Refrigeration Machine Based}} on {{Four Quantum Dots}}},
  author = {Venturelli, Davide and Fazio, Rosario and Giovannetti, Vittorio},
  year = {2013},
  month = jun,
  journal = {Phys. Rev. Lett.},
  volume = {110},
  number = {25},
  pages = {256801},
  issn = {0031-9007, 1079-7114},
  doi = {10.1103/PhysRevLett.110.256801},
  urldate = {2025-05-30},
  copyright = {http://link.aps.org/licenses/aps-default-license},
  langid = {english}
}

@article{czartowski2024,
  title = {Catalytic Transformations for Thermal Operations},
  author = {Czartowski, Jakub and De Oliveira Junior, A.},
  year = {2024},
  month = aug,
  journal = {Phys. Rev. Res.},
  volume = {6},
  number = {3},
  pages = {033203},
  issn = {2643-1564},
  doi = {10.1103/PhysRevResearch.6.033203},
  urldate = {2025-05-31},
  langid = {english}
}

@article{chen2019,
  title = {Birkhoff-von {{Neumann}} Theorem and Decomposition for Doubly Stochastic Tensors},
  author = {Chen, Haibin and Qi, Liqun and Caccetta, Louis and Zhou, Guanglu},
  year = {2019},
  month = dec,
  journal = {Linear Algebra Appl.},
  volume = {583},
  pages = {119--133},
  issn = {00243795},
  doi = {10.1016/j.laa.2019.08.027}
}

@book{sagawa2022,
  title = {Entropy, {{Divergence}}, and {{Majorization}} in {{Classical}} and {{Quantum Thermodynamics}}},
  author = {Sagawa, Takahiro},
  year = {2022},
  series = {{{SpringerBriefs}} in {{Mathematical Physics}}},
  volume = {16},
  publisher = {Springer Singapore},
  address = {Singapore},
  doi = {10.1007/978-981-16-6644-5},
  urldate = {2024-11-19},
  copyright = {https://www.springer.com/tdm},
  isbn = {978-981-16-6643-8 978-981-16-6644-5}
}

@article{Cangemi2024,
  title = {Quantum Engines and Refrigerators},
  author = {Cangemi, Loris Maria and Bhadra, Chitrak and Levy, Amikam},
  year = {2024},
  month = oct,
  journal = {Phys. Rep.},
  volume = {1087},
  pages = {1--71},
  issn = {03701573},
  doi = {10.1016/j.physrep.2024.07.001},
  urldate = {2025-06-04},
  langid = {english}
}

@article{kosloff2014,
  title = {Quantum {{Heat Engines}} and {{Refrigerators}}: {{Continuous Devices}}},
  author = {Kosloff, Ronnie and Levy, Amikam},
  year = {2014},
  month = apr,
  journal = {Ann. Rev. Phys. Chem.},
  volume = {65},
  number = {1},
  pages = {365--393},
  issn = {0066-426X, 1545-1593},
  doi = {10.1146/annurev-physchem-040513-103724},
  urldate = {2025-06-04}
}

@article{abah2012,
  title = {Single-{{Ion Heat Engine}} at {{Maximum Power}}},
  author = {Abah, O. and Ro{\ss}nagel, J. and Jacob, G. and Deffner, S. and {Schmidt-Kaler}, F. and Singer, K. and Lutz, E.},
  year = {2012},
  month = nov,
  journal = {Phys. Rev. Lett.},
  volume = {109},
  number = {20},
  pages = {203006},
  issn = {0031-9007, 1079-7114},
  doi = {10.1103/PhysRevLett.109.203006},
  urldate = {2025-09-05},
  copyright = {http://link.aps.org/licenses/aps-default-license},
  langid = {english}
}

@article{barker2022,
  title = {Experimental {{Verification}} of the {{Work Fluctuation-Dissipation Relation}} for {{Information-to-Work Conversion}}},
  author = {Barker, David and Scandi, Matteo and Lehmann, Sebastian and Thelander, Claes and Dick, Kimberly A. and {Perarnau-Llobet}, Mart{\'i} and Maisi, Ville F.},
  year = {2022},
  month = jan,
  journal = {Phys. Rev. Lett.},
  volume = {128},
  number = {4},
  pages = {040602},
  issn = {0031-9007, 1079-7114},
  doi = {10.1103/PhysRevLett.128.040602},
  urldate = {2025-09-05},
  langid = {english}
}

@article{camati2016,
  title = {Experimental {{Rectification}} of {{Entropy Production}} by {{Maxwell}}'s {{Demon}} in a {{Quantum System}}},
  author = {Camati, Patrice A. and Peterson, John P. S. and Batalh{\~a}o, Tiago B. and Micadei, Kaonan and Souza, Alexandre M. and Sarthour, Roberto S. and Oliveira, Ivan S. and Serra, Roberto M.},
  year = {2016},
  month = dec,
  journal = {Phys. Rev. Lett.},
  volume = {117},
  number = {24},
  pages = {240502},
  issn = {0031-9007, 1079-7114},
  doi = {10.1103/PhysRevLett.117.240502},
  urldate = {2025-09-05},
  copyright = {http://link.aps.org/licenses/aps-default-license}
}

@article{masuyama2018,
  title = {Information-to-Work Conversion by {{Maxwell}}'s Demon in a Superconducting Circuit Quantum Electrodynamical System},
  author = {Masuyama, Y. and Funo, K. and Murashita, Y. and Noguchi, A. and Kono, S. and Tabuchi, Y. and Yamazaki, R. and Ueda, M. and Nakamura, Y.},
  year = {2018},
  month = mar,
  journal = {Nat. Commun.},
  volume = {9},
  number = {1},
  pages = {1291},
  issn = {2041-1723},
  doi = {10.1038/s41467-018-03686-y},
  urldate = {2025-09-05}
}

@article{zhou2023,
  title = {Full {{Tunability}} and {{Quantum Coherent Dynamics}} of a {{Driven Multilevel System}}},
  author = {Zhou, Yuan and Gu, Sisi and Wang, Ke and Cao, Gang and Hu, Xuedong and Gong, Ming and Li, Hai-Ou and Guo, Guo-Ping},
  year = {2023},
  month = apr,
  journal = {Phys. Rev. Appl.},
  volume = {19},
  number = {4},
  pages = {044053},
  issn = {2331-7019},
  doi = {10.1103/PhysRevApplied.19.044053},
  urldate = {2025-09-08}
}

@phdthesis{biswas2023,
  author       = {Tanmoy Biswas},
  title        = {Finite size effects in quantum thermodynamics},
  school       = {University of Gdańsk, Poland},
  year         = {2023},
  type         = {PhD Thesis},
  supervisor   = {Michal Horodecki},
  cosupervisor = {Kamil Korzekwa}}

@book{horn2012,
  title = {Matrix Analysis},
  author = {Horn, Roger A. and Johnson, Charles R.},
  year = {2012},
  edition = {2nd ed},
  publisher = {Cambridge University Press},
  address = {Cambridge ; New York},
  isbn = {978-0-521-83940-2},
  langid = {english},
  lccn = {QA188 .H66 2012}
}

@article{linden2010,
  title = {How {{Small Can Thermal Machines Be}}? {{The Smallest Possible Refrigerator}}},
  shorttitle = {How {{Small Can Thermal Machines Be}}?},
  author = {Linden, Noah and Popescu, Sandu and Skrzypczyk, Paul},
  year = {2010},
  month = sep,
  journal = {Phys. Rev. Lett.},
  volume = {105},
  number = {13},
  pages = {130401},
  issn = {0031-9007, 1079-7114},
  doi = {10.1103/PhysRevLett.105.130401},
  urldate = {2025-09-09},
  copyright = {http://link.aps.org/licenses/aps-default-license}
}

@article{aamir2025a,
  title = {Thermally Driven Quantum Refrigerator Autonomously Resets a Superconducting Qubit},
  author = {Aamir, Mohammed Ali and Jamet Suria, Paul and Mar{\'i}n Guzm{\'a}n, Jos{\'e} Antonio and {Castillo-Moreno}, Claudia and Epstein, Jeffrey M. and Yunger Halpern, Nicole and Gasparinetti, Simone},
  year = {2025},
  month = feb,
  journal = {Nat. Phys.},
  volume = {21},
  number = {2},
  pages = {318--323},
  issn = {1745-2473, 1745-2481},
  doi = {10.1038/s41567-024-02708-5}
}

@article{yu2019,
  title = {Quantum Self-Contained Refrigerator in Terms of the Cavity Quantum Electrodynamics in the Weak Internal-Coupling Regime},
  author = {Yu, C. S. and Guo, B. Q. and Liu, T.},
  year = {2019},
  month = mar,
  journal = {Opt. Express},
  volume = {27},
  number = {5},
  pages = {6863},
  issn = {1094-4087},
  doi = {10.1364/OE.27.006863}
}

@article{koukoulekidis2021,
  title = {The Geometry of Passivity for Quantum Systems and a Novel Elementary Derivation of the {{Gibbs}} State},
  author = {Koukoulekidis, Nikolaos and Alexander, Rhea and Hebdige, Thomas and Jennings, David},
  year = {2021},
  month = mar,
  journal = {Quantum},
  volume = {5},
  pages = {411},
  issn = {2521-327X},
  doi = {10.22331/q-2021-03-15-411}
}

@article{scharlau2018,
  title = {Quantum {{Horn}}'s Lemma, Finite Heat Baths, and the Third Law of Thermodynamics},
  author = {Scharlau, Jakob and Mueller, Markus P.},
  year = {2018},
  month = feb,
  journal = {Quantum},
  volume = {2},
  pages = {54},
  issn = {2521-327X},
  doi = {10.22331/q-2018-02-22-54},
  urldate = {2025-09-10}
}

@article{deoliveirajunior2025,
  title = {Heat as a {{Witness}} of {{Quantum Properties}}},
  author = {De Oliveira Junior, A. and Brask, Jonatan Bohr and {Lipka-Bartosik}, Patryk},
  year = {2025},
  month = feb,
  journal = {Phys. Rev. Lett.},
  volume = {134},
  number = {5},
  pages = {050401},
  issn = {0031-9007, 1079-7114},
  doi = {10.1103/PhysRevLett.134.050401}
}

@article{junior2024,
  title = {Quantum Catalysis in Cavity Quantum Electrodynamics},
  author = {Junior, A. De Oliveira and {Perarnau-Llobet}, Mart{\'i} and Brunner, Nicolas and {Lipka-Bartosik}, Patryk},
  year = {2024},
  month = may,
  journal = {Phys. Rev. Res.},
  volume = {6},
  number = {2},
  pages = {023127},
  issn = {2643-1564},
  doi = {10.1103/PhysRevResearch.6.023127}
}
\end{document}